\documentclass[fleqn,usenatbib]{mnras}
\usepackage{newtxtext,newtxmath}
\usepackage{comment}
\usepackage{multirow}
\usepackage{mathtools}
\usepackage[T1]{fontenc}
\usepackage{ae,aecompl}

\usepackage{graphicx}	
\usepackage{amsmath}	

\usepackage{amssymb}	
\usepackage[ruled,vlined]{algorithm2e}
\usepackage{algpseudocode}
\usepackage{enumitem} 
\usepackage{xfrac} 
\usepackage{comment} 

\setlist[itemize]{leftmargin=*}

\newcommand{\norm}[1]{\|#1\|}

\newcommand{\ud}{\mathrm{d}}

\newcommand{\vect}[1]{\boldsymbol{#1}}

\newcommand{\derfrac}[2]{\frac{\ud #1}{\ud #2}}

\newcommand{\msol}[1]{{#1}\:\mathrm{M_\odot}}
\newcommand{\Sigmasol}[1]{{#1}\:\mathrm{M_\odot\:pc^{-2}}}

\newcommand{\mstar}{\texttt{MSTAR}}

\newcommand{\ketju}{\texttt{KETJU}}
\newcommand{\gthree}{\texttt{GADGET-3}}
\newcommand{\gfour}{\texttt{GADGET-4}}

\newcommand{\sersic}{S\'ersic}

\newcommand{\boxtube}{\ensuremath{N_\mathrm{box}/N_\mathrm{tube}}}

\title[Evolution of red nuggets]{{The supermassive black hole merger driven evolution of high-redshift red nuggets into present-day cored early-type galaxies}}
\author[A. Rantala et al.]{Antti Rantala$^{1}$\thanks{E-mail: anttiran@mpa-garching.mpg.de}, Alexander Rawlings$^{2}$, Thorsten Naab$^{1}$, Jens Thomas$^{3}$, \newauthor Peter H. Johansson$^{2}$\\
$^{1}$Max-Planck-Institut f\"ur Astrophysik, Karl-Schwarzschild-Str. 1, 
D-85748, Garching, Germany\\
$^{2}$Department of Physics, University of Helsinki, P.O. Box 64 (Gustaf Hällströmin katu 2), FI-00014, University of Helsinki, Finland\\
$^{3}$ Max-Planck-Institut f\"ur Extraterrestriche Physik, Giessenbach-Str. 1, D-85748, Garching, Germany
}
\date{Accepted XXX. Received YYY; in original form ZZZ}
\pubyear{2024}

\begin{document}
\label{firstpage}
\pagerange{\pageref{firstpage}--\pageref{lastpage}}
\maketitle

\begin{abstract}
Very compact ($R_\mathrm{e}\lesssim1$ kpc) massive quiescent galaxies (red nuggets) are more abundant in the high-redshift Universe ($z\sim2$--$3$) than today. Their size evolution can be explained by collisionless dynamical processes in galaxy mergers which, however, fail to reproduce the diffuse low-density central cores in the local massive early-type galaxies (ETGs). We use sequences of major and minor merger N-body simulations starting with compact spherical and disk-like progenitor models to investigate the impact of supermassive black holes (SMBHs) on the evolution of the galaxies. With the \ketju{} code we accurately follow the collisional interaction of the SMBHs with the nearby stellar population and the collisionless evolution of the galaxies and their dark matter halos. We show that only models including SMBHs can simultaneously explain the formation of low-density cores up to sizes of $R_\mathrm{b} \sim 1.3$ kpc with mass deficits in the observed range and the rapid half-mass size evolution. In addition, the orbital structure in the core region (tangentially biased orbits) is consistent with observation-based results for local cored ETGs. The displacement of stars by the SMBHs boost the half-mass size evolution by up to a factor of two and even fast rotating progenitors (compact quiescent disks) lose their rotational support after $6$--$8$ mergers. We conclude that the presence of SMBHs is required for merger driven evolution models of high redshift red nuggets into local ETGs. 
\end{abstract}



\section{Introduction}
Cosmological simulations supported by observations have revealed that the formation of massive early-type galaxies can be divided into two distinct phases (e.g  \citealt{Naab2009,Feldmann2010,Oser2010,Oser2012,Johansson2012,Hilz2012,Hilz2013,Wellons2015,RodriguezGomez2016,Qu2017,Naab2017,Ene2020,Cannarozzo2023}). At early times ($z\gtrsim 1.5$) galaxies grow rapidly by in-situ star formation fueled by cold gas flows and accretion of multiple star-bursting progenitors. The latter extended phase, characterized by mass and especially size growth of the galaxies is primarily driven by the accretion of stars originally formed in smaller satellite galaxies. In addition, simple arguments based on the virial theorem can be used to show that multiple minor mergers will produce a more extended galaxy compared to accreting the same stellar mass in few major mergers \citep{Naab2009,Bezanson2009}.

The effective radii $R_\mathrm{e}$ of massive early-type galaxies (ETGs) have been observed to grow by a factor of $\sim 4$--$7$ since $z\sim2$ (e.g. \citealt{Daddi2005,Trujillo2007,vanDokkum2008,Buitrago2008,Franx2008,Bezanson2009,Cenarro2009,vanDokkum2009,Carrasco2010,vanDokkum2010,Cassata2011,vanderWel2011,Szomoru2011,Szomoru2012,vanderWel2014,Belli2014}) driven mainly by accretion of gas-poor satellites \citep{vanDokkum2010,vanDokkum2015}, as expected in the two-phased formation scenario. The $z\sim2$ compact, quiescent early-type galaxies, sometimes termed `red nuggets', have typically large stellar masses of the order of $M_\star \sim \msol{10^{11}}$ while having small effective radii, $R_\mathrm{e}\lesssim1\:\mathrm{kpc}$.

The exact surface brightness profile shapes and rotation properties of red nuggets remain debated. While the median S\'ersic indices $n$ of $z=2$ quiescent galaxies are comparable to the local massive ETGs with $n_\mathrm{median} = 3.7$ \citep{Szomoru2012}, there is also evidence for surface brightness profiles that are closer to exponential. Selecting galaxies at a constant number density of $n=2\times10^{-4}\:\mathrm{Mpc}^{-3}$, \cite{vanDokkum2010} found that massive galaxies become more exponential at high redshifts, e.g. their mean S\'ersic index decreases from $n\sim6$ at $z=0$ down to $n\sim2$ at $z=2$. Individual massive, quiescent, rotationally supported disk galaxies above $z=2$ have also been discovered, as reported by \cite{Toft2017} who found a disky (axis ratio $0.59\substack{+0.03\\-0.09}$) rapidly rotating system at $z=2.1478$ with $\log_\mathrm{10}(M_\star/\mathrm{M}_\odot) = 11.15\substack{+0.23\\-0.23}$ and $R_\mathrm{e} = 1.73\substack{+0.34\\-0.27}\ \rm kpc$. The  S\'ersic index of the galaxy $n=1.01\substack{+0.12\\-0.06}$, is consistent with an exponential system. The shape of the system is congruous with observations that show that compact massive galaxies tend to have highly elliptic shapes both at high redshift \citep{vanderWel2011,Buitrago2013} and at $z\sim0.15$ \citep{Trujillo2012}.

Due to the stochastic nature of mergers in the hierarchical picture of cosmological structure formation, it is possible for a subset of $z\sim 2$ massive compact galaxies to only experience a few mergers during their later evolution, and thus remain relatively compact until the present day \citep{Quilis2013}. Such massive, compact galaxies with uniformly old stellar populations are termed local relic galaxies. Local relic candidates have been identified with varying degrees of relicness quantified by their star formation and accretion histories (e.g. \citealt{Trujillo2009,Taylor2010,FerreMateu2012,Trujillo2012,Poggianti2013,vandenBosch2015,Saulder2015,Tortora2016}). 
Local ($z\lesssim0.1$) relic galaxies are relatively rare, with \cite{Quilis2013} estimating that fewer than $0.1\%$ of massive galaxies have accreted less than $10\%$ of their stars after their formation, corresponding to a number density of $n \sim10^{-6}\: \mathrm{Mpc}^{-3}$ \citep{Trujillo2014}. 

Despite their low expected numbers, local relics have been confirmed, with NGC 1277 \citep{Trujillo2014}, Mrk 1216 and PGC 032873 \citep{FerreMateu2017} being the prime examples. The three local relic galaxies have stellar masses ranging between $\msol{1.2\times10^{11}} \leq M_\star \leq \msol{2.9\times10^{11}}$ and effective radii in the range of $1.2 \ \mathrm{kpc} \leq R_\mathrm{e} \leq 2.3 \ \mathrm{kpc}$. The stellar masses and effective radii with their uncertainties are displayed in Table \ref{table: local-relics}.
\begin{table}
    \centering
    \begin{tabular}{lccc}
        \hline
        local relic & $M_\star$ & $R_\mathrm{e}$ & reference\\
        galaxy    & [$\msol{10^{11}}$] & [kpc] & \\
        \hline
        NGC 1277 & $1.2\pm0.4$ & $1.2\pm0.1$ & \citet{Trujillo2014}\\
        Mrk 1216 & $2.0\pm0.8$ & $2.3\pm0.1$ & \citet{FerreMateu2017}\\
        PGC 032873 & $2.3\pm0.9$ & $1.8\pm0.2$ & \citet{FerreMateu2017}\\
        \hline
    \end{tabular}
    \caption{The stellar masses $M_\star$ and effective radii $R_\mathrm{e}$ of three local relic galaxies NGC 1277, Mrk 1216 and PGC 032873. The galaxies are described in more detail in the text.}
    \label{table: local-relics}
\end{table}
Resembling the $z\sim2$ red nuggets, NGC 1277, Mrk 1216, and PGC 032873 show elongated, disky shapes, strong rotation and highly peaked velocity dispersion profiles \citep{Trujillo2014, FerreMateu2017}. A straightforward comparison to the S\'ersic profiles of red nuggets is challenging as multiple S\'ersic components are required for good fits of the surface brightness profiles for both NGC 1277 and Mrk 1216 \citep{Yildrim2015}.
In addition to local relics, various survey programs have also revealed the existence of several tens of relic candidates at intermediate redshifts \citep[$0.1 \lesssim z \lesssim 1.0$, e.g.][]{Valentinuzzi2010,Damjanov2014,Spiniello2021,Spiniello2024}.

The most massive present-day ($z\sim0$) ETGs differ considerably from $z\sim2$ compact red quiescent galaxies and the local relics in their structural and kinematic properties. The brightest ($M_\mathrm{B} \lesssim -20.5$) and most massive ($M_\star \gtrsim \msol{10^{11}}$) early-type galaxies observed in the local Universe are characterized by their extended sizes, shallow central surface brightness profiles and slow rotation (e.g. \citealt{Kormendy1996,Faber1997,Quenneville2024}). Massive galaxies also ubiquitously host supermassive black holes in their nuclei (e.g. \citealt{Kormendy1995,Ferrarese2005,Kormendy2013}). Many host galaxy properties have been found to tightly correlate with the mass of their SMBHs, including the bulge mass ($M_\bullet$-$M_\star$ relation; \citealt{Häring2004}) and the velocity dispersion ($M_\bullet$-$\sigma$ relation; \citealt{Ferrarese2000,Gebhardt2000,Tremaine2002}) suggesting a co-evolution of massive galaxies and their SMBHs. 

Binaries of SMBHs formed in the aftermath of galaxy mergers have been suggested to explain the observed faint, diffuse central cores of massive early-type galaxies, and there are indications of core-SMBH relations linking the SMBH mass to the amount of missing starlight in the core region (mass deficit; \citealt{Graham2004,deRuiter2005,Lauer2007,Kormendy2009}) as well as the spatial extent of the cores \citep{Rusli2013,Thomas2016}. The characteristic kinematic fingerprint of the SMBH binary core scouring, the tangentially biased velocity anisotropy ($\beta<0$) in the core region, has also been observed in massive, cored early-type galaxies \citep{Thomas2014}.

Theoretical arguments and later detailed numerical simulations have established a picture of the formation, evolution, and mergers of SMBH binaries in the aftermath of mergers of massive gas-poor galaxies (e.g. \citealt{Begelman1980,Ebisuzaki1991,Quinlan1996,Quinlan1997,Milosavljevic2001,Milosavljevic2003,Merritt2005,Merritt2006,Merritt2013,Nasim2021}). First, dynamical friction \citep{Chandrasekhar1943} from stars, gas and dark matter causes the two SMBHs to sink to the center of the merger remnant, and the SMBHs form a gravitationally bound binary. The binary subsequently interacts with the stars in the galactic nucleus, losing energy to the surrounding stars in complex three-body interactions while shrinking towards smaller separations. Only stars with a sufficiently small amount of angular momentum (the so-called loss-cone population) can interact with the SMBH binary, and thus the binary preferentially ejects stars from radial orbits, leading to a slow build-up of stars on circular orbits in the core region (e.g. \citealt{Thomas2014}). The rapidly lowered central stellar density and the more slowly developed tangentially biased velocity dispersion of the core can be viewed as the two phases of the core formation, transforming the orbit structure of the stars in the nucleus \citep{Frigo2021}

The so-called final-parsec problem, i.e. the stalling of the SMBH binary at parsec-scale separations due to insufficient loss-cone refilling, seems to be avoided in realistic simulation models of axisymmetric, triaxial or rotating nuclei (e.g. \citealt{Berczik2006,Preto2011,Khan2011,Khan2013,HolleyBockelmann2015,Khan2016,Gualandris2017,Rantala2017,Mirza2017}), although questions related to the potentially insufficient mass resolution of the current simulation models still remain \citep{Vasiliev2014}. In gas-rich environments both nuclear star formation and sub-pc scale gas disk dynamics affect the SMBH binary evolution \citep{Liao2023}. Detailed hydrodynamical simulations including SMBH binary dynamics highlight the importance of SMBH accretion and feedback for calculating the binary coalescence time-scales \citep{Liao2024a,Liao2024b}. Finally, at centiparsec separations, relativistic radiation-reaction effects become important \citep{Poisson2014}, and the binary radiates its remaining energy and angular momentum \citep{Peters1963,Peters1964} until merging into a single SMBH in a burst of gravitational radiation.

Despite the general success of the theoretical and numerical models explaining the evolution of compact quiescent galaxies at $z\sim2$ into the present-day population of massive ETGs, open questions still remain (see \citealt{Naab2017} for a review). 
Large cosmological box runs and zoom-in simulations that probe the size growth and rotation of ETGs from high redshift to the present-day typically ignore the small-scale SMBH binary dynamics (and thus core formation) due to the instant merger procedure at large separations of the order of $\sim 0.1$ kpc (see e.g. the discussion of the merger procedure in \citealt{Johansson2009}).

Recently \cite{Mannerkoski2021,Mannerkoski2022} performed the first cosmological zoom-in simulations that include accurate, non-softened dynamics with post-Newtonian equations of motion for SMBH interactions using the \gthree-based regularized tree code \ketju{} \citep{Rantala2017}. While the code can in principle follow the SMBHs and SMBH binary dynamics including core scouring in cosmologically evolving early-type galaxies, the accurate SMBH dynamics requires that the mass ratio between the SMBHs and stellar population particles is large enough ($M_\bullet/m_{\star}\gtrsim 100$). This corresponds to accurate small-scale SMBH dynamics at redshifts $z\lesssim1$ in these particular simulations. However, improved versions of the code and upcoming high performance computing hardware capabilities will ameliorate the problem in the future  allowing for accurate SMBH binary dynamics for a wider range in redshift (Keitaanranta et al., in prep.). 

The main question to be answered in this study is whether the $z=2$ compact early-type galaxies can be the progenitors of present-day massive, cored and slowly rotating early-type galaxies. We especially focus on the question whether the sizes, cores and rotation properties can be simultaneously explained by a single physical process, namely major and minor galaxy mergers including SMBHs. As simulations in a full cosmological setting with accurate SMBH dynamics below $z\lesssim2$ is still beyond the current computational capabilities, we choose here instead to run samples of controlled, idealized gas-free galaxy mergers.

The article is structured as follows. After the introduction we review the \gfour-based simulation code \ketju{} and the initial conditions for our numerical simulations in Section \ref{section: methods-ics}. We present the results on galaxy size growth and SMBH binary core scouring in Section \ref{section: sizegrowth} and Section \ref{section: coresection}. Next, in Section \ref{section: scouring-global} we describe how core scouring can affect the global properties of ETGs, especially for large mass deficits. Finally, we discuss the shape, velocity anisotropy and kinematic structure of our simulated galaxies in Section \ref{section: shape-rotation} and conclude in Section \ref{section: conclusions}.

\section{Methods and numerical simulations}\label{section: methods-ics}

\subsection{Simulation code}

\subsubsection{The \ketju{} code: summary}

We use the recent publicly available\footnote{\href{https://www.mv.helsinki.fi/home/phjohans/ketju}{https://www.mv.helsinki.fi/home/phjohans/ketju}} implementation \citep{Mannerkoski2023} of the \ketju{} simulation code \citep{Rantala2017}. The novel code version is itself based on the tree / fast multipole method (FMM) gravity solver of \texttt{GADGET-4} \citep{Springel2021}. The main functionality of \ketju{} is to add an algorithmically regularized region around each SMBH in the simulation in which the gravitational dynamics can be accurately calculated without the need to employ gravitational softening. The size of the regularized regions typically ranges from a few parsecs up to tens of parsecs chosen based on the stellar density around each SMBH so that the maximum number of stellar particles within any regularized region does not exceed $N_\star\sim 5000$--$10000$. The spatial size of the regularized region places an upper limit for the softening length $\epsilon = \epsilon_\star = \epsilon_\bullet$ of SMBHs and stellar particles in \texttt{GADGET-4}. This is because \ketju{} requires $r_\mathrm{region}>2.8 \times \epsilon$ in order to have all SMBH-SMBH and star-SMBH interactions non-softened. In this work only SMBH and star particles participate in the regularization integration while all interactions including dark matter (DM) particles are always softened. Star-star interactions within the regularized regions are softened as well.

\subsubsection{\mstar: the \ketju{} integrator}
The algorithmically regularized (e.g. \citealt{Mikkola1999,Preto1999}) integrator of \ketju{} is based on the \mstar{} integrator of \cite{Rantala2020}. \mstar{} utilizes three main ingredients to achieve accurate and efficient orbital integration. First, the equations of motion of the particles in the regularized regions are time-transformed. Together with the use of the standard leapfrog integration, the time transformation avoids the issues of the diverging Newtonian potential at small separations \citep{Mikkola1999,Mikkola2002,Mikkola2006,Mikkola2008}. Next, \mstar{} uses a specialized minimum spanning tree (MST) coordinate system to reduce the numerical floating-point round-off error \citep{Rantala2020}. Finally, in order to reach the user-desired integration accuracy, \mstar{} employs the Gragg-Bulirsch-Stoer (GBS) extrapolation method \citep{Gragg1965, Bulirsch1966}. The computational cost of the GBS method is mitigated by efficient two-fold MPI parallelization of the code, based on the fact that the different sub-step divisions (e.g. \citealt{Deuflhard1983}) required by the method are independent of each other \citep{Rantala2020}.

Compared to the version of \cite{Rantala2020}, the \mstar{} version (the \ketju{} integrator) included in the public \ketju{} code \citep{Mannerkoski2023} includes a dynamical order control for the GBS extrapolation procedure \citep{Hairer1993} and an optional star-star softening. In addition, the code includes post-Newtonian (PN) equations of motion up to order PN3.5 with optional spin terms for the SMBH-SMBH interactions \citep{Thorne1985, Blanchet2014}, and finally a prescription for the relativistic gravitational wave (GW) recoil kicks at SMBH-SMBH mergers \citep{Zlochower2015}. For additional PN and GW implementation details, see \cite{Mannerkoski2023}.

\subsubsection{\gfour: galactic-scale dynamics}
Outside the regularized regions, the gravitational dynamics in \ketju{} is treated as in the standard \gfour{} code. Compared to the previous \ketju{} implementations \citep{Rantala2017} based on \gthree{} (e.g. \citealt{Springel2005}), the \gfour{} \citep{Springel2021} based \ketju{} \citep{Mannerkoski2023} has several advantages. First, \gfour{} has an option to use the so-called hierarchical integration scheme based on hierarchical Hamiltonian splitting instead of the common block time-step scheme. The hierarchical scheme is more efficient than the block step scheme for systems with extended time-step hierarchies as the rapidly evolving parts of the simulation decouple from the more slowly evolving regions. Importantly, hierarchical integration allows for manifest momentum conservation if coupled with a pair-wise force calculation method such as direct summation or the FMM, which is not possible using the traditional block time-step scheme.

\gfour{} includes an accurate fast multipole method (FMM) gravity solver (\citealt{Greengard1987,Dehnen2000,Dehnen2002,Dehnen2014,Zhu2020}) in addition to the traditional one-sided tree code \citep{Appel1985,Barnes1986,Makino1993,Springel2001,Springel2005}. The FMM force calculations between particles (and tree nodes) are always pair-wise and the corresponding particle kicks are always synchronous, together with the hierarchical integration making the force calculation and time integration algorithm manifestly momentum conserving. This could in principle be very important for SMBH binary hardening phases in \ketju, as spurious loss-cone filling originating from one-sided tree force errors might potentially affect the SMBH binary hardening rates, and thus their merger time-scales \citep{Gualandris2017}. However, as shown by \cite{Mannerkoski2023} in their Fig. 8 the loss cone filling rates are nearly identical when using the one-sided tree compared to the FMM solver.

\subsubsection{Simulation code parameters}

We use the following options, user-given accuracy and error tolerance parameters for the \gfour-\ketju{} code. For \gfour, we use hierarchical integration and the FMM gravity solver with second-order multipoles. The \gfour{} force and integration error tolerances are set to $\alpha = 0.005$ and $\eta = 0.002$. For the \mstar{} parameters, the Gragg-Bulirsch-Stoer per-step relative error tolerance parameter is set to $\epsilon_\mathrm{GBS} = 10^{-7}$ and the output time relative tolerance parameter is $\epsilon_\mathrm{t} = 10^{-3}$. Post-Newtonian terms are included in the SMBH-SMBH interactions up to PN3.5.

The gravitational interactions of the dark matter particles are softened using a Plummer-equivalent softening length of $\epsilon_\mathrm{dm} = 65$ pc. The stellar component has a softening length of $\epsilon_\star = 6.5$ pc. The size of the regularized regions around the SMBHs is set to $r_\mathrm{region} = 20$ pc, fulfilling the necessary criterion of $r_\mathrm{region} > 2.8 \times \epsilon_\star$ \citep{Rantala2017}, resulting in typically a few hundred stellar particles in the regions. Star-star softening is used in the regularized regions \citep{Mannerkoski2021,Mannerkoski2023} as our stellar population particles have masses $\gg\msol{1}$. SMBH-SMBH and SMBH-star interactions are always unsoftened. In this study the interactions between the SMBHs and the collisionless dark matter particles are always softened, and dark matter particles are not included in the regularized integration. We note that a scenario in which SMBHs and dark matter can strongly scatter each other has been recently explored by \cite{Partmann2023} using \gthree-\ketju.

\subsection{Initial conditions for galaxy models}

\subsubsection{Isolated galaxy models: spherical and disky stellar components}

We generate isolated multi-component early-type galaxy initial conditions (ICs) later to be merged with each other in our merger simulation setups. We use a modified version of the standard galaxy IC generation code \texttt{MakeDiskGalaxy} \citep{Springel2005diskmerger}, which constructs equilibrium galaxy models consisting of a dark matter halo, a stellar component (an axisymmetric thick disk or a spherical bulge) and a central SMBH. We construct three types of isolated models: two host galaxy progenitor models (spherical, disky) and lower-mass spherical satellite galaxy models. 

The host galaxy stellar mass is set to $M_\mathrm{\star,host} = \msol{10^{11}}$ while the satellites have $M_\mathrm{\star,sat} = \msol{2\times10^{10}}$, resulting in a satellite-host mass ratio of $M_\mathrm{\star,sat}/M_\mathrm{\star,host} = 1/5$. The host galaxies have initially a three-dimensional stellar half-mass radius of $r_\mathrm{h} = 1 \ \mathrm{kpc}$. We construct satellites with five different half-mass radii ranging from $r_\mathrm{h} = 0.97$ kpc to $r_\mathrm{h} = 3.36$ kpc, following the observed mass-size relation for early-type galaxies \citep{vanderWel2014} between redshifts $z=2$ and $z=0$. In minor mergers beyond generation five (generations 6--8), we use the fifth ($z=0$) satellite model. The satellite models are used in merger simulation sequences described in section \ref{section: sequences}. The basic physical properties of the isolated IC galaxy models are listed in Table \ref{table: ic-isolated}. The stellar components of the galaxy models are realized using $N_\mathrm{\star} = 10^5$ per host galaxy and $N_\mathrm{\star} = 2\times 10^4$ for the satellites. The necessary condition \citep{Rantala2017} for well resolved stellar component for the SMBH-star interactions in \ketju{} ($M_\bullet/m_\star \gtrsim 100$ with $m_\star = M_\star/N_\star$) is always fulfilled as $\min{(M_\bullet)}/m_\star = 400$ in the simulations of this study.

The host galaxy stellar components are either spherical (S) or disky\footnote{We note that `diskiness' in this study refers to the thick, axisymmetric shape of the stellar component and not to the traditional disky versus boxy isophotal shape classification of ETGs.} (D). The satellites galaxy models are always spherical. The spherical stellar components follow the Hernquist profile \citep{Hernquist1990} with a central power-law density slope $(\rho(r) \propto r^{-\gamma})$ of $\gamma=1$. The Hernquist stellar models are isotropic and non-rotating. The disky models are rotationally supported and have exponential surface density profiles. The axisymmetric disky models are thick in the vertical direction with the minor-to-major axis ratio being $c/a = 0.5$ within $r_\mathrm{h}$. 

\subsubsection{Dark matter halos}

The dark matter halos of the galaxies follow the Hernquist profile, just as the spherical stellar components. The host halo mass of $M_\mathrm{dm} = \msol{10^{12}}$ is chosen following the simulation setup of \cite{Hilz2012,Hilz2013}. In general, it is thought that dark matter may play a minor role in the central parts of high-redshift massive galaxies due to their early dissipative evolution (e.g. \citealt{Toft2012}). In addition NGC 1277, a local relic galaxy is known to be dark matter deficient \citep{Comeron2023}. Motivated by such arguments, the Hernquist scale radii of the host halos are chosen so that the dark matter fraction $f_\mathrm{dm} = M_\mathrm{dm}/(M_\mathrm{dm} + M_\star)$ is low, only $f_\mathrm{dm} = 3\%$ within $1 \times r_\mathrm{h}$ and $f_\mathrm{dm} = 21\%$ within $5 \times r_\mathrm{h}$. The satellite galaxy models always have $f_\mathrm{dm} = 20\%$ within the half-mass radius, more typical of present-day massive early-type galaxies \citep{Cappellari2013,Courteau2015}. The host galaxy halos are realized with $N_\mathrm{dm} = 10^6$ particles while the satellite halos consist of $N_\mathrm{dm} = 2\times10^5$ particles.

\subsubsection{SMBH properties}

We construct two versions of each galaxy model, with and without a central SMBH. The SMBHs are placed at rest at the center of the isolated galaxy model. The host galaxy SMBH mass is set to $M_\mathrm{\bullet,host} = \msol{2\times10^9}$ following the SMBH mass estimates of \cite{Graham2016} for the local relic galaxy NGC 1277. The satellite SMBH mass is set to $M_\mathrm{\bullet,sat} = \msol{4\times10^8}$ corresponding to the satellite-host black hole mass ratio of $1/5$.

\begin{centering}
\begin{table*}
\begin{tabular}{|c|c|c|c|c|c|c|c|c|} 
\hline
 model & shape & $\rho(r)$, $\Sigma(r)$ & rotation & $M_\mathrm{dm}$ & $M_\mathrm{\star}$ & $M_\mathrm{\bullet}$ & $r_\mathrm{h,\star} $ & reference $z$\\
 & & & & [$10^{11}\mathrm{M_\odot}$] & [$10^{11}\mathrm{M_\odot}$] & [$10^{9}\mathrm{M_\odot}$] & [kpc] &\\
\hline
 host-S & spherical & Hernquist & $\times$ & $10.0$ & $1.0$ & 2.0 & $1.00$ & $2.00$\\ 
 host-D & axisymmetric & exponential & $\checkmark$ & $10.0$ & $1.0$ & 2.0 & $1.00$ & $2.00$\\ 
 \hline
 sat-1 & spherical & Hernquist & $\times$ & $2.0$ & $0.2$ & 0.4 & $0.97$ & $2.00$\\
 sat-2 & spherical & Hernquist & $\times$ & $2.0$ & $0.2$ & 0.4 & $1.25$ & $1.56$\\ 
 sat-3 & spherical & Hernquist & $\times$ & $2.0$ & $0.2$ & 0.4 & $1.68$ & $1.13$\\ 
 sat-4 & spherical & Hernquist & $\times$ & $2.0$ & $0.2$ & 0.4 & $2.48$ & $0.69$\\ 
 sat-5 & spherical & Hernquist & $\times$ & $2.0$ & $0.2$ & 0.4 & $3.36$ & $0.25$\\ 
 \hline
\end{tabular}
\caption{The initial galaxy models of this study. The redshift $z$ is for the reference of the mass-size relation as discussed in the main text. The disky shape refers to an axisymmetric profile with axis ratios $b/a=1.0$ and $c/a=0.5$ with an exponential stellar surface density profile. Two versions of each profile are generated, one with a central SMBH and another without.}
\label{table: ic-isolated}
\end{table*}
\end{centering}

\subsection{Simulation sequences}\label{section: sequences}

\subsubsection{Merger orbits}

We adopt the basic galaxy merger orbit configuration from the previous \ketju{} ETG merger studies \citep{Rantala2017,Rantala2018,Rantala2019}. The initial separation of the nuclei of the galaxies is set to $r_\mathrm{sep} = 30 \ \mathrm{kpc}$. The two merging galaxies are placed on a nearly parabolic orbit \citep{Khochfar2006,Naab2006} with the initial relative velocity scaled by a factor of $f_\mathrm{vel} = M(<r_\mathrm{sep})/M_\mathrm{tot}$, ensuring a relatively rapid galaxy merger and limiting the computational expense of the simulations. For all the minor merger simulations in this study the initial pericenter distance $r_\mathrm{p}$ is fixed to $r_\mathrm{p}=2$ kpc. 

For the major merger orbits of the spherical models we use two different initial pericenter distances ($r_\mathrm{p,lo} = 1$ kpc and $r_\mathrm{p,hi} = 9$ kpc) resulting in a slowly rotating and a fast-rotating major merger remnant in the case of the spherical progenitors. Somewhat counter-intuitively for the spherical non-rotating progenitors, the major merger orbit with $r_\mathrm{p,lo}=1$ kpc results in the fast-rotating merger remnant and $r_\mathrm{p,hi}=9$ kpc in a slowly rotating remnant, even though the former orbit has less angular momentum than the latter. The reason for this lies in how the stellar orbital angular momentum is re-distributed during the merger and how much of it is still remaining when the stellar bulges merge. The process is explained in more detail in Appendix \ref{section: appendix-rotation}. For the major merger simulations between disky models we use $r_\mathrm{p,mi} = 3$ kpc.

\subsubsection{Merger sequences: overview}

\begin{figure*}
\includegraphics[width=\textwidth]{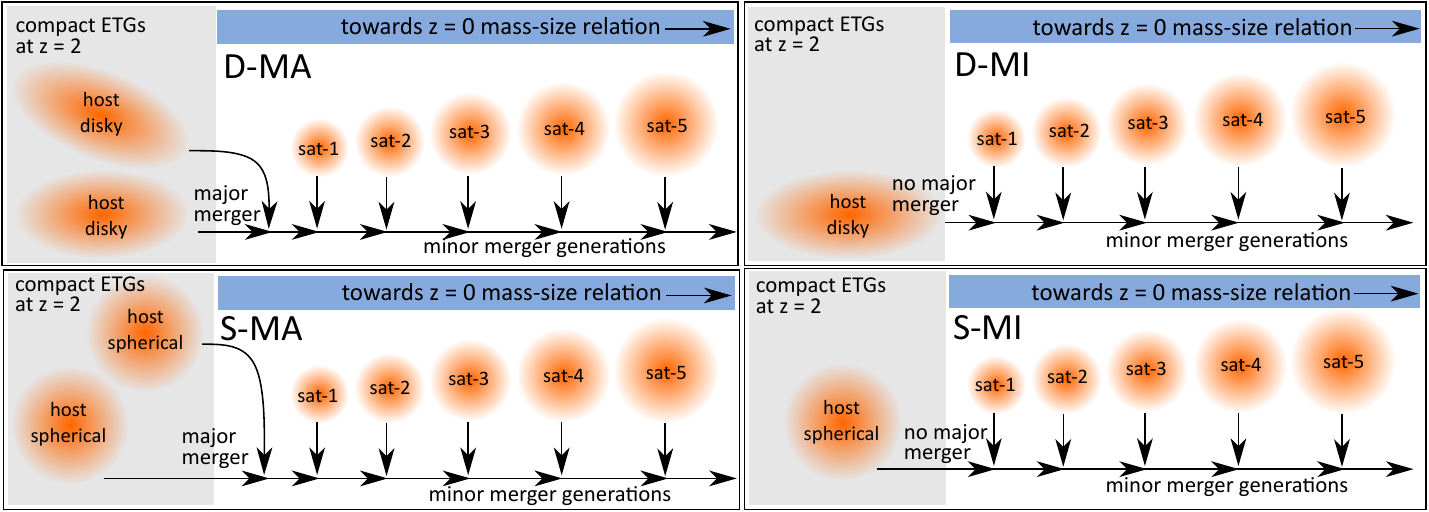}
\caption{The four basic types of merger sequences performed for this study. The sat-5 satellite model is used in minor merger generations beyond generation five. Top left: the D-MA sequence. Two disky ETGs are first merged, after which minor mergers are repeated until the remnant reaches the $z=0$ mass-size relation for ETGs. Top right: the D-MI sequence. A disky progenitor experiences a sequence of minor mergers without the initial major merger. Bottom panels: the merger sequences S-MA and S-MI. The two merger sequences correspond to the sequences on the top row, but now the progenitor galaxies are spherical and non-rotating. This study includes in total ten merger sequences. Each of the sequences D-MA, D-MI, S-MA and S-MI is run twice, with and without SMBHs. In addition, the setup S-MA is repeated with an alternative merger orbit, both with and without SMBHs, resulting in a fast-rotating progenitor after the major merger and before the minor merger sequence. These two sequences are labeled S-MA-SROT and S-MA-FROT depending on the rotation after the major merger. Note that the sizes of the galaxies are just for illustrative purposes and do not reflect the relative masses or spatial extents of the stellar components.}
\label{fig: mergers-inkscape}
\end{figure*}

We construct in total $10$ different merger sequences from the host galaxies and satellites described in the previous section and in Table \ref{table: ic-isolated}. We mainly focus on varying the following simulation parameters. First, we study merger sequences with different progenitor galaxy shapes and rotation properties, i.e. whether the host galaxy is spherical (S) or disky (D). Next, we explore the effect of SMBHs by running each merger sequence with (label BH) and without central SMBHs (no label). Finally, we investigate how including a single major merger (MA) or alternatively excluding it (MI) in the merger sequence affects the evolution of the merger remnant. If a major merger is included in the merger sequence, the sequence begins with it, followed by the multiple minor mergers. 

These three options (disky/spherical host, major merger or not, SMBHs or not) yield $8$ different basic merger sequences, outlined in Fig. \ref{fig: mergers-inkscape}. The last $2$ merger sequences explore different galaxy major merger orbit configurations leading to very different merger remnant rotation properties (slow or fast rotation) from the same progenitors.

The satellite galaxies are merged into the host progenitors from isotropic random directions but in such a manner that any Nth minor merger occurs from the same direction in all the merger sequences. The minor mergers are continued until the merger remnant galaxies reach the $z=0$ mass-size ($M_\star$-$R_\mathrm{e}$) relation for early-type galaxies. The required size growth to reach the observed relation is substantial as the effective radius of the host galaxy progenitors is only $R_\mathrm{e}\sim 0.75 \ \mathrm{kpc}$, and a size growth by a factor of $\sim5$--$10$ is needed to reach the relation \citep{Shen2003}. We run $0$ or $1$ major mergers and $7$--$8$ minor mergers for each ten merger sequences. In merger generations $6$--$8$ we use the satellite model sat-5.

In this work, we study exclusively isolated binary galaxy mergers with and without SMBHs. During hierarchical structure formation, multiple SMBHs commonly end up in the same halo, which may lead to complex interactions (e.g. \citealt{Mannerkoski2021}). We also note that the effect of having multiple massive black holes in a single galactic nucleus has been recently explored by \cite{Partmann2023}.

\subsubsection{Merger sequences with disky progenitors}

\begin{table}
\centering
\begin{tabular}{|l|c|c|c|} 
\hline
sequence & host & major merger & SMBH\\
\hline
D-MA-BH & disky & $\checkmark$ & $\checkmark$\\
D-MA & disky & $\checkmark$ & $\times$\\
\hline
D-MI-BH & disky & $\times$ & $\checkmark$\\
D-MI & disky & $\times$ & $\times$\\
\hline
\end{tabular}
\caption{The four merger sequences from disky progenitors (D-MA-BH, D-MA, D-MI-BH, D-MI) with ($\checkmark$) and without ($\times$) a major merger and with and without central SMBHs.}
\label{table: sequence-disky}
\end{table}

The four merger sequences with a disky progenitor galaxy (D-MA-BH, D-MA, D-MI-BH, D-MI) are listed in Table \ref{table: sequence-disky}. The sequences D-MA-BH and D-MA begin with a major galaxy merger while the sequences D-MI-BH and D-MI do not include a major merger. Central SMBHs are included in sequences D-MA-BH and D-MI-BH. In the disky major merger simulations one galaxy lies initially in the orbital plane while the other is tilted by $\theta=45^\circ$.

\subsubsection{Merger sequences with spherical progenitors}

\begin{centering}
\begin{table}
\begin{tabular}{|l|c|c|c|c|} 
\hline
sequence & host & major & rotation after & SMBH\\
 &  & merger & major merger & \\
\hline
S-MA-SROT-BH & spherical & $\checkmark$ & slow & $\checkmark$\\
S-MA-SROT & spherical & $\checkmark$ & slow & $\times$\\
\hline
S-MA-FROT-BH & spherical & $\checkmark$ & fast & $\checkmark$\\
S-MA-FROT & spherical & $\checkmark$ & fast & $\times$\\
\hline
S-MI-BH & spherical & $\times$ & & $\checkmark$\\
S-MI & spherical & $\times$ & & $\times$\\
\hline
\end{tabular}
\caption{The six merger sequences beginning from spherical progenitors.}
\label{table: sequence-spherical}
\end{table}
\end{centering}

The mergers beginning from spherical progenitors are listed in Table \ref{table: sequence-spherical}. There are $6$ merger sequences in total. In addition to the sequences with and without a major merger and SMBHs we perform the major merger simulations with two different merger orbit configurations by varying the initial pericenter separation $r_\mathrm{p}$. The two major merger orbit configurations result in a slow rotating merger remnant and a fast rotating merger remnant. The minor merger sequences from these two progenitors are labeled as S-MA-SROT-BH, S-MA-SROT (slow rotation, with and without SMBHs) and S-MA-FROT-BH, S-MA-FROT (fast rotation, with and without SMBHs). The two sequences without a major merger are S-MI-BH and S-MI depending if SMBHs is present in the minor mergers or not.

\subsubsection{SMBH binary eccentricities}

\begin{figure}
\includegraphics[width=\columnwidth]{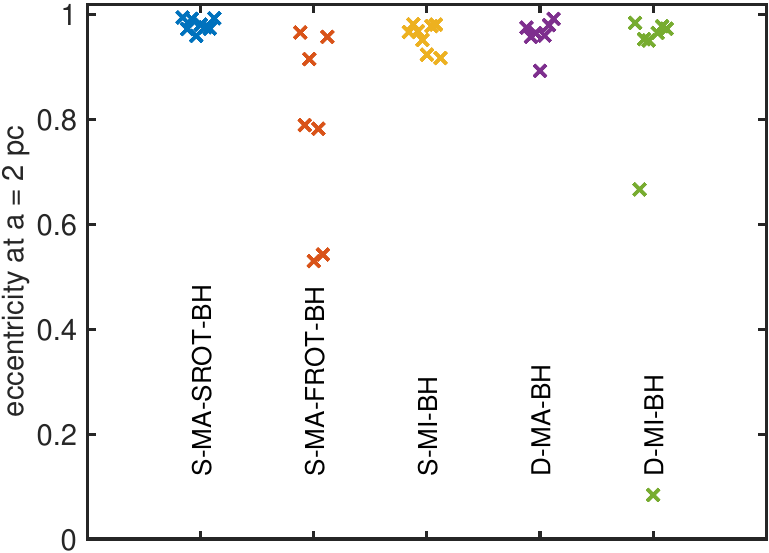}
\caption{The eccentricities of the SMBH binaries in the simulation sequences at a binary semi-major separation of $a=2$ pc. At this separation the gravitational wave emission has not yet circularized the binary orbit. Most of the binaries are very eccentric with $e\gtrsim0.9$. A number of less eccentric binaries occur in the samples S-MA-FROT-BH and D-MI-BH.}
\label{fig: eccentricity}
\end{figure}

The SMBH binary eccentricities in the minor merger simulations at binary semi-major axis of $a=2$ pc are shown in Fig. \ref{fig: eccentricity}. The particular semi-major axis was chosen so that even the most eccentric SMBH binaries in our simulations have not yet entered the dynamical regime dominated by gravitational wave emission.

Most of the SMBH binaries in the minor merger simulations are highly eccentric with $e>0.9$, especially in the runs S-MA-SROT-BH, S-MI-BH and D-MA-BH. The simulation sequence S-MA-FROT-BH includes four moderately eccentric binaries with $0.5<e<0.8$ while the sequence D-MI-BH involves two binaries with low and moderate eccentricities with $e=0.08$ and $e=0.67$. The in general high binary eccentricities likely somewhat depend on the selected galaxy merger orbits, although there is recent evidence that stochastic processes may play a significant role in determining the initial SMBH binary eccentricities \citep{Rawlings2023}.

\section{The size growth of massive early-type galaxies via major and minor mergers}\label{section: sizegrowth}

\subsection{Reaching the $z=0$ mass-size relation}\label{section: reach-mass-size-relation}

\begin{figure*}
\includegraphics[width=\textwidth]{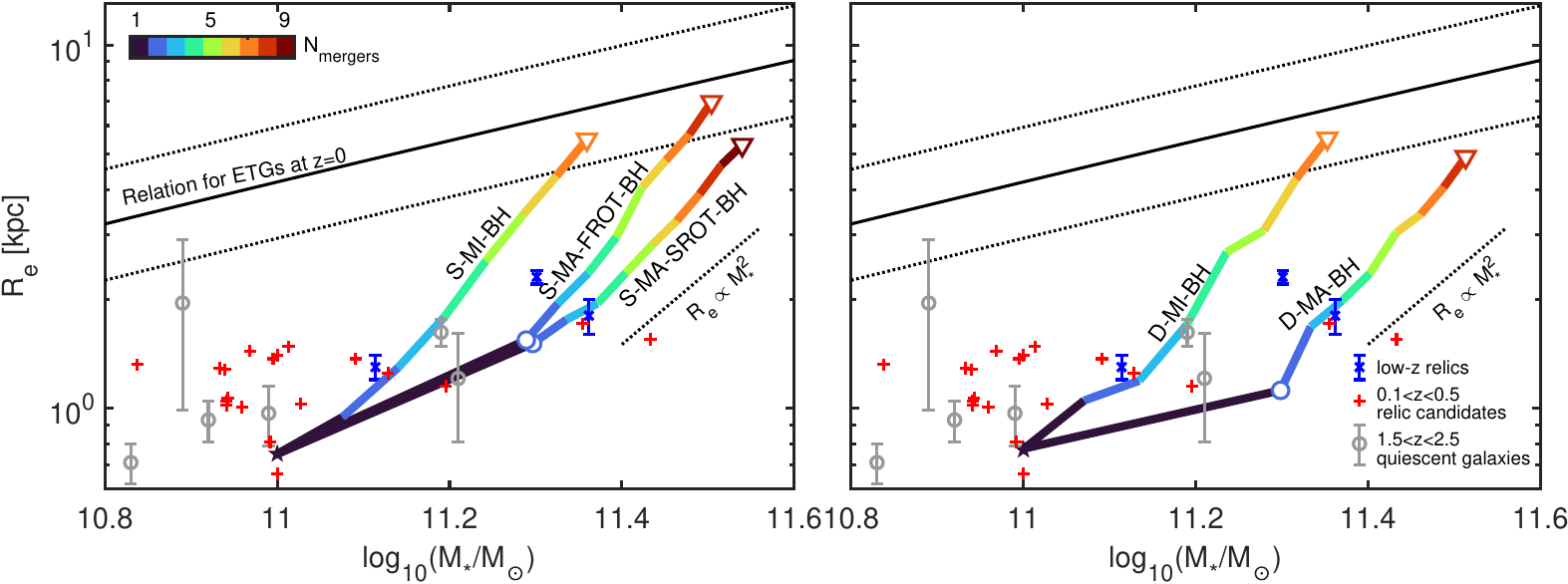}
\caption{The evolution of the mass-size relation in the simulation sequences including SMBH: S-MA-SROT-BH, S-MA-FROT-BH and S-MI-BH (left panel) and D-MA-BH with D-MI-BH in the right panel. Both panels also show observational comparison data from local red nugget like relics NGC 1277, PGC 032873 and Mrk 1216 \citep{Trujillo2014,FerreMateu2017}, intermediate-redshift ($0.1<z<0.5$) relic candidates \citep{Spiniello2021,Spiniello2024} and high redshift ($1.5<z<2.5$) quiescent (sSFR<$0.3/t_\mathrm{Hubble}$) galaxies \citep{Szomoru2013}. The local SDSS $z=0$ mass-size relation for ETGs \citep{Shen2003} is displayed for comparison purposes. In general, $5$--$8$ minor mergers are required to reach the $z=0$ relation starting from a compact progenitor galaxy. Major mergers result in size growth $R_\mathrm{e}\propto M_\star^{\alpha_\bullet}$ with $\alpha_\bullet\sim1$, almost parallel to the relation, so more minor mergers are required to reach the local mass-size relation in the merger samples beginning with a major merger. For minor mergers with SMBHs the size growth is steeper with $\alpha_\bullet>2.3$. This is further discussed in the text.}
\label{fig: mass_size_bh}
\end{figure*}

\begin{figure*}
\includegraphics[width=\textwidth]{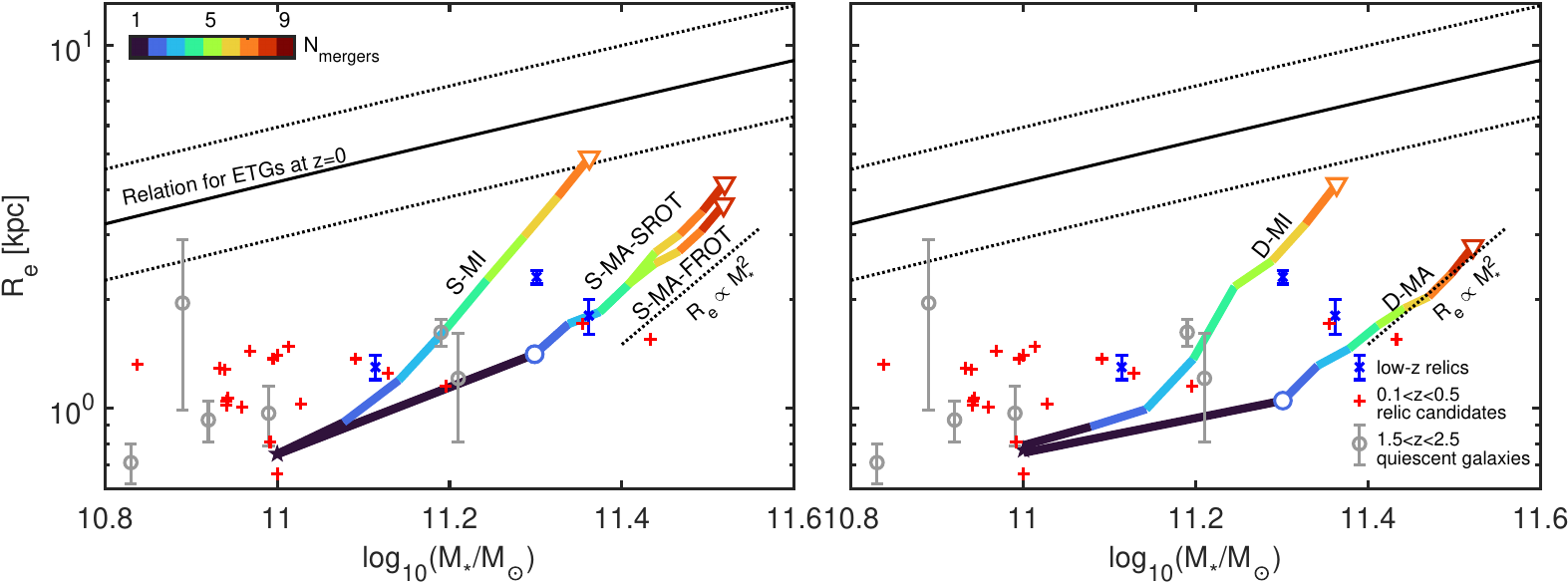}
\caption{The evolution of the mass-size relation as in Fig. \ref{fig: mass_size_bh} but for simulation samples without SMBHs: S-MA-SROT, S-MA-FROT and S-MI (left) and D-MA with D-MI (right). The size growth in the merger sequences is less prominent than in the corresponding simulations with SMBHs.}
\label{fig: mass_size_nobh}
\end{figure*}

\begin{table}
    \begin{tabular}{lcccc}
        sequence & major+SMBH & major & minor+SMBH & minor\\
        \hline
        S-MA-SROT & 1.02 & 0.92 & 2.31 & 1.80\\
        S-MA-FROT & 1.08 & 0.92 & 3.10 & 2.10\\
        S-MI & - & - & 2.51 & 2.33 \\
        D-MA & 0.54 & 0.48 & 2.81 & 1.79 \\
        D-MI & - & - & 2.46 & 2.10 \\
        \hline
    \end{tabular}
    \caption{The power-law exponents $\alpha$ of the mass-size relation $R_\mathrm{e} \propto M_\star^\alpha$ for the major and minor merger sequences.}
    \label{table: hilz-alpha}
\end{table}

First we focus on the evolution of the mass-size ($M_\star$-$R_\mathrm{e}$) relation of the initially compact early-type galaxies in the merger sequences. The evolution of the mass-size relation in the consecutive merger simulations is presented in Fig. \ref{fig: mass_size_bh} (including SMBHs) and in Fig. \ref{fig: mass_size_nobh} (without SMBHs). Both our spherical and disky progenitors at $M_\star=\msol{10^{11}}$ have initial effective radii of $R_\mathrm{e} = 0.75$ kpc. Compared to the local SDSS ETG mass-size relation \citep{Shen2003} at the same mass, the progenitors are by a factor of $\sim5.6$ smaller in spatial size.

In our simulations the initially compact galaxy models can reach the $z=0$ mass-size relation after $6$--$8$ generations of minor mergers. As already expected from simple virial arguments \citep{Naab2009,Bezanson2009}, minor mergers are more efficient than major mergers in increasing the galaxy sizes. A major merger moves a galaxy in the mass-size plane in an almost parallel direction to the $z=0$ relation, thus requiring more minor mergers in the sequences beginning with a major merger to reach the relation. 

The power-law size growth of galaxies in merger sequences is typically characterized as $R_\mathrm{e} \propto M_\star^\alpha$ with $\alpha>2$ required to explain the observed size growth of ETGs since $z=2$ \citep{vanDokkum2010}.  The values of the galaxy size growth power-law index $\alpha$ from our mergers sequences are collected in Table \ref{table: hilz-alpha}. In our simulations, equal-mass major mergers of spherical progenitors show an almost linear trend in size growth with $0.92 \lesssim \alpha \lesssim 1.08$, consistent with the results in the literature \citep{Boylan-Kolchin2005,Ciotti2007,Bezanson2009,Nipoti2009,Hilz2012,Hilz2013}. Mergers including SMBHs result in somewhat larger values for $\alpha$ with $\alpha=0.92$ (without SMBHs) and $1.02 \lesssim \alpha_\bullet \lesssim 1.08$ (with SMBHs). For the disky progenitors the size growth in a major merger is more limited with $0.48 \lesssim \alpha \lesssim 0.54$.

Studying our minor merger simulations, the power-law exponent $\alpha$ of the galaxy mass-size relation is close to $\alpha \sim 1.8$ in merger setups S-MA-SROT and D-MA, and $\alpha\gtrsim2.1$ in all other minor merger setups, consistent with the observed size growth of ETGs since $z=2$ \citep{vanDokkum2010}. The size growth with 
$\alpha=2.33$ in the spherical minor merger sample without SMBHs S-MI is very close to the results of \cite{Hilz2013} who used a similar setup to ours, although with less compact progenitor galaxies. 

Including SMBHs in the merger sequences increases $\alpha$ in all minor merger simulations. The effect is especially prominent in the samples S-MA-SROT ($\alpha_\bullet-\alpha=0.51$), S-MA-FROT ($\alpha_\bullet-\alpha=1.0$) and D-MA ($\alpha_\bullet-\alpha=1.02$), each of which begins with a major galaxy merger. For the simulations without a major merger $ \alpha_\bullet-\alpha=0.18$ for the merger sequence S-MI and $ \alpha_\bullet-\alpha=0.36$ for the simulations D-MI. We explore this phenomenon in Section \ref{section: scouring-rh}, where we argue that the reason for the additional size growth is simply the displacement of stellar material due to core scouring by high-mass SMBHs.

\subsection{Comparison to observed compact ETGs}

Next, we compare our simulations results to the observed masses and sizes of three local relic galaxies (NGC 1277, PGC 032873 and Mrk 1216; \citealt{Trujillo2014,FerreMateu2017}), relic candidates at $0.1<z<0.5$ \citep{Spiniello2021,Spiniello2024} and massive quiescent\footnote{We define a galaxy as quiescent if it has a specific star formation rate sSFR$<0.3 / t_\mathrm{Hubble}$.} galaxies at $1.5<z<2.5$ \citep{Szomoru2013}. The observed masses and radii of the galaxies are displayed alongside our simulation results in the mass-size plane in Fig. \ref{fig: mass_size_bh} and Fig. \ref{fig: mass_size_nobh}.

Our initial isolated galaxy models have masses ($M_\star=\msol{10^{11}}$) comparable both to the high-redshift quiescent galaxies and the intermediate-redshift relic candidates while having somewhat smaller effective radii than most of the observed compact galaxies. In the simulation samples without major mergers, the location of the simulated galaxies in the mass-size plane remains consistent with the observed high and intermediate redshift galaxies during the first $N_\mathrm{merg}=2$-$3$ minor mergers. After $N_\mathrm{merg} \geq3$ the galaxies experiencing minor mergers have larger effective radii $R_\mathrm{e}$ than the observed galaxies in the samples. Three of our models with two minor mergers (S-MI-BH, D-MI-BH and S-MI) are consistent with the stellar mass and effective radius of the local relic NGC 1277. Compared to the two other local relics, PGC 032873 and Mrk 1216, our minor merger models without a major merger have in general larger sizes for their stellar mass.

The merger samples beginning with a major merger already have larger stellar masses than any of the galaxies in the high-redshift quiescent galaxy sample after the major merger. Nevertheless, they have effective radii of $1.05 \ \mathrm{kpc} \lesssim R_\mathrm{e} \lesssim 1.54 \ \mathrm{kpc}$ comparable to the most massive galaxies in the high-redshift quiescent sample, $1.2 \ \mathrm{kpc} \lesssim R_\mathrm{e} \lesssim 1.6 \ \mathrm{kpc}$. Comparing to the intermediate redshift relic candidates, the simulated galaxies in the samples including a major merger have two observational counterparts with masses close to $\log_\mathrm{10}(M_\star/\mathrm{M}_\odot)=11.4$. The simulated galaxies have effective radii comparable to these galaxies $(1.55 \ \mathrm{kpc} \lesssim R_\mathrm{e} \lesssim 1.71 \ \mathrm{kpc})$ after a major merger and $N_\mathrm{merg}=1$-$3$ minor mergers, depending on the simulation sample. The local relic PGC 032873 has also a similar stellar mass $\log_\mathrm{10}(M_\star/\mathrm{M}_\odot)=11.36$ and effective radius ($R_\mathrm{e}=1.8$ kpc) as our simulated galaxies after a major merger and $N_\mathrm{merg}=2$ minor mergers. The third local relic, Mrk 1216 lies between the major and minor merger simulation samples in the mass-size plane. These results highlight the possibility that local relic galaxies may have experienced a few mergers, but not the full sequence to evolve into the largest ETGs in the local Universe.

\section{Evolution of galactic nuclei in major and minor mergers}\label{section: coresection}

\subsection{Surface density profiles}

\begin{figure*}
\includegraphics[width=0.8\textwidth]{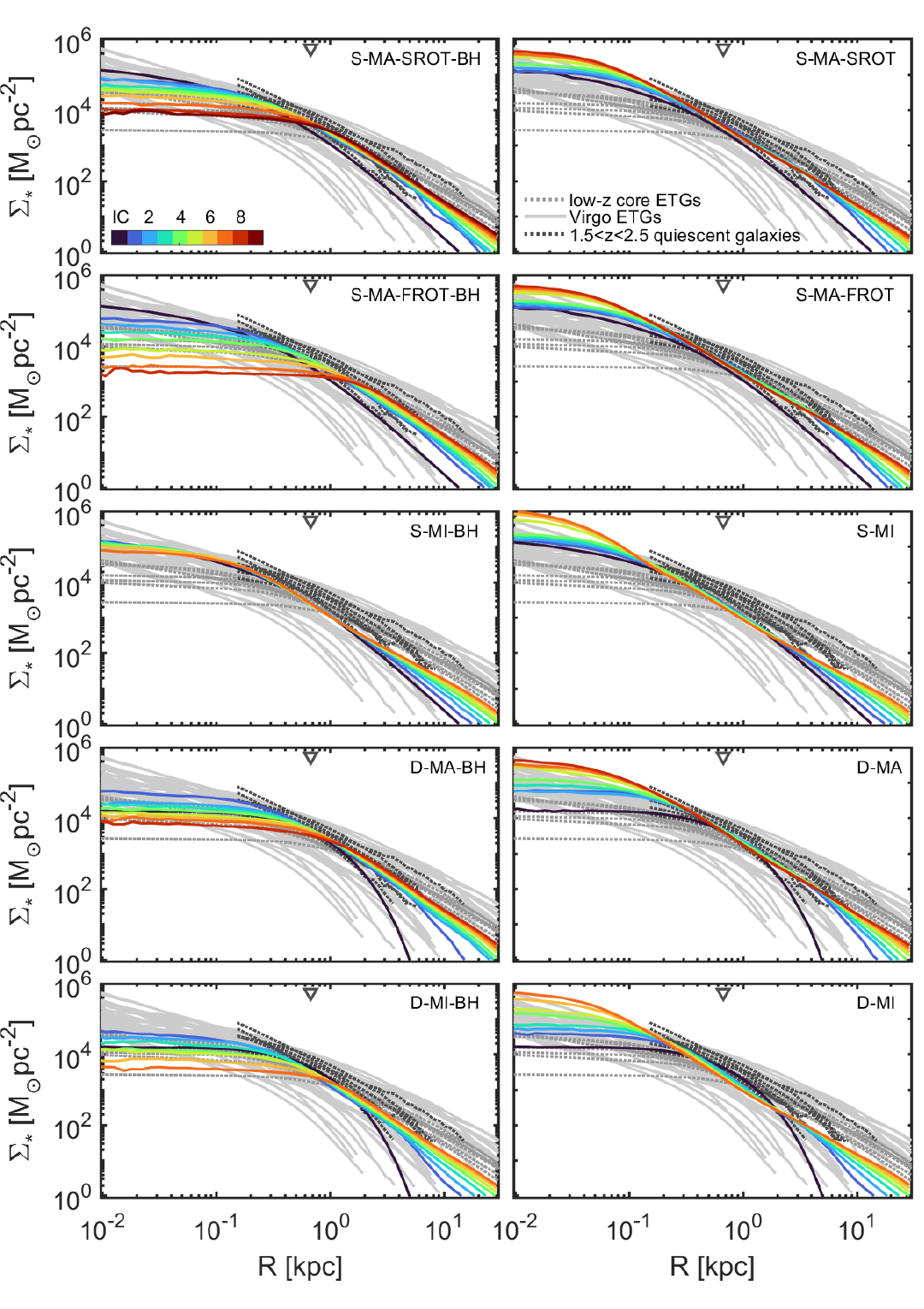}
\caption{The azimuthally averaged radial stellar surface density profiles $\Sigma_\star(R)$ of the merger remnants in the simulation samples including SMBHs (left) column and without SMBHs (right column). Each profile is estimated immediately after the SMBH merger in the simulation as discussed in the main text and is} averaged over $N_\mathrm{los}=50$ random viewing orientations. The number of mergers increases from $N_\mathrm{merg}=1$ to $N_\mathrm{merg}=7$--$9$ as indicated in the legend. For comparison, we show the surface density profiles of selected local core ETGs \citep{Rusli2013}, typical Virgo ETGs \citep{Kormendy2009} and $1.5<z<2.5$ quiescent galaxies \citep{Szomoru2013}. The central parts of the high-redshift quiescent galaxies are not resolved at radii below $r\lesssim 0.7$ kpc \citep{Szomoru2012}, indicated by a triangle symbol at the top of each panel. Only models with SMBHs (excluding S-MI-BH which is further discussed in the text) resemble the surface density profiles of local, massive, cored ETGs after the merger sequences while without SMBHs the final surface density profiles are mildly cuspy at the centers with values for central $\Sigma_\star(R)$, both from spherical and disky initial conditions.
\label{fig: surfacedensity}
\end{figure*}

We show the stellar surface density profiles of our simulated galaxy sample after each galaxy merger in Fig. \ref{fig: surfacedensity}. The profiles are shown at the time of the SMBH mergers in the simulations, typically $0.3$ Gyr--$3.5$ Gyr since the beginning of each run, depending on the central stellar density and the eccentricity of the SMBH binary at its formation. Each surface density profile is averaged over $N_\mathrm{los}=50$ random line-of-sight viewing directions. The profiles are azimuthally (circularly) averaged and range from $R=0.01$ kpc in the inner parts to $R=30$ kpc in the outer parts. The surface density in the outer parts ($R\gtrsim1$ kpc) increases with each merger generation in the all simulation samples.

For comparison to our simulation results, we show the surface density profiles of high redshift ($1.5<z<2.5$) quiescent galaxies \citep{Szomoru2013}, local ETGs in the Virgo cluster with masses comparable or higher than for observed $z=2$ compact ETGs \citep{Kormendy2009,Szomoru2012} and a number of local massive cored ETGs \citep{Rusli2013}. The central parts of the high-redshift quiescent galaxies are not resolved at radii below $r\lesssim 0.7 \ \rm kpc$ \citep{Szomoru2012}, and their surface density profiles are consistent both with our initial compact spherical setups and the disky setups after a single major or minor merger.

\subsubsection{Spherical models including a major merger}

The spherical initial galaxy model host-S has a central (at $R=0.01$ kpc) stellar surface density of $\Sigma_\star \sim \Sigmasol{1.3\times10^5}$. In the spherical setups beginning with a major merger (S-MA-SROT-BH and S-MA-FROT-BH) the central surface density monotonically decreases with each merger generation as the SMBH binaries scour increasingly extended, flat low-density cores into the galaxies. 

SMBH binaries in the S-MA-FROT-BH sample have on average lower eccentricities when they form compared to their counterparts with slowly rotating host galaxies S-MA-SROT-BH as shown in Fig. \ref{fig: eccentricity}. Thus, they enter the regime dominated by GW losses at smaller spatial separations, having lost comparably more orbital energy in the three-body slingshot hardening phase, resulting in larger and less dense cores. After one major merger and eight minor mergers, the central surface density of the S-MA-FROT-BH setup has decreased by two orders of magnitude.

On the other hand, in the spherical sequences with a major merger but without SMBHs, the central surface density gradually increases with each merger generation, as opposed to their counterparts which include SMBHs. This is due to violent relaxation (e.g. \citealt{LyndenBell1967,Hilz2012,Hilz2013}). The setups S-MA-SROT and S-MA-FROT have very similar mildly cuspy central structure after the minor mergers with $\Sigma_\star \sim \Sigmasol{5\times10^5}$ at $R=0.01$ kpc.

The final merger remnants of the simulation sequences S-MA-SROT-BH and S-MA-FROT-BH are in very good agreement with the observed local massive ETGs (e.g. \citealt{Rusli2013}) with an extended, flat low-density core. Without SMBHs the merger remnants have too high central surface densities compared to the observed core ETGs and instead rather resemble cuspy power-law galaxies without cores \citep{Ferrarese1994}.

\subsubsection{Spherical models with only minor mergers}

As shown in Fig. \ref{fig: eccentricity}, in the spherical sequence S-MI-BH including only minor mergers the formed SMBH binaries typically always have high eccentricities ($e\gtrsim0.9$) and thus merge rapidly driven by gravitational wave emission. Interestingly, the central surface density profile remains very close to the initial mildly cuspy profiles despite the SMBH mergers as only relatively little energy is transferred from the binary to the stellar component. We performed additional SMBH merger simulations in which the SMBH binaries initially have circular orbits in the setup. In these runs, presented in Appendix \ref{section: appendix-sims-gwecc}, a low-density core was scoured.

Despite the relatively unaffected central surface density profile $\Sigma_\star(R)$ the comparison of the simulation sample S-MI-BH to its counterpart S-MI without SMBHs clearly demonstrates that SMBH core scouring has an effect on the profile. Without SMBHs, the central surface density increases by each merger, reaching $\Sigma_\star \sim \Sigmasol{1\times10^6}$ at $R=0.01$ kpc after the sequence of seven minor mergers. Thus, in the sample S-MI-BH the SMBHs do not scour a flat, extended low-density core, but rather prevent the central surface density from increasing by a factor of $\sim7$ during the minor merger sequence. As we define the mass deficit as the missing central mass in the runs with SMBHs compared to their counterparts without SMBHs, it is possible to have $M_\mathrm{def}>0$ even if the central $\Sigma_\star(R)$ would remain unaffected.

Compared to local ETGs, the models in the sample S-MI-BH lack a flat, low-density core, even though their central surface densities are lower than in the corresponding sample without SMBHs. Simulated galaxy merger remnants in both sequences S-MA-BH and S-MI resemble mildly cuspy ETGs with their Hernquist-like centers.

\subsubsection{Disky models}

The disky progenitor model host-D has a central ($R=0.01$ pc) surface density of $\Sigma_\star \sim \Sigmasol{1.9\times10^4}$, a factor of $\sim7.5$ lower compared to the spherical progenitor host-S. In the sequences D-MA and D-MI without SMBHs, the central surface density rapidly increases via minor mergers as the in-falling spherical satellites with a Hernquist density profile fill the initially close to flat exponential center of the disky model. After the merger sequences, the central surface densities have reached values of $\Sigma_\star \sim \Sigmasol{4.4\times10^5}$ (D-MA) and $\Sigma_\star \sim \Sigmasol{5.5\times10^5}$ (D-MI).

In the disky models including SMBHs, the central surface density initially increases in the first few mergers, both in the sequences with (D-MA-BH) and without (D-MI-BH) a major merger. As with the model S-MI, SMBH core scouring still acts in the simulations as the central surface densities without SMBHs are systematically higher. As the later satellites are less bound (larger $r_\mathrm{h}$) than the early ones as described in Table \ref{table: ic-isolated}, core scouring can lead to increasingly lowered central density in later mergers. The surface density of the disky progenitor is again reached after $N_\mathrm{merg} \sim4$-$5$ minor mergers, and the final central surface densities after the entire merger sample are $\Sigma_\star \sim \Sigmasol{9.6\times10^3}$ and $\Sigma_\star \sim \Sigmasol{4.4\times10^3}$ for the samples D-MA-BH and D-MI-BH, respectively. The final central $\Sigma_\star$ is very similar in the merger samples D-MI-BH and S-MA-FROT-BH despite their different (spherical or disky) progenitor properties.

The initial outer surface density profile of the exponential disky progenitor is somewhat steep compared to the \sersic{} outer parts of observed ETGs. A single minor or major merger is enough to transform the outer simulated profile into a \sersic-like mildly curving shape. In the inner parts, only models with SMBHs resemble the flat, low-density central parts of the local, massive cored ETGs. Simulated merger remnants without SMBHs on the other hand resemble cuspy power-law galaxies due to violent relaxation, even though the slope of the central stellar cusp is not very steep.

\subsection{Mass deficits with and without SMBHs}\label{section: Mdef}

\begin{figure}
\includegraphics[width=\columnwidth]{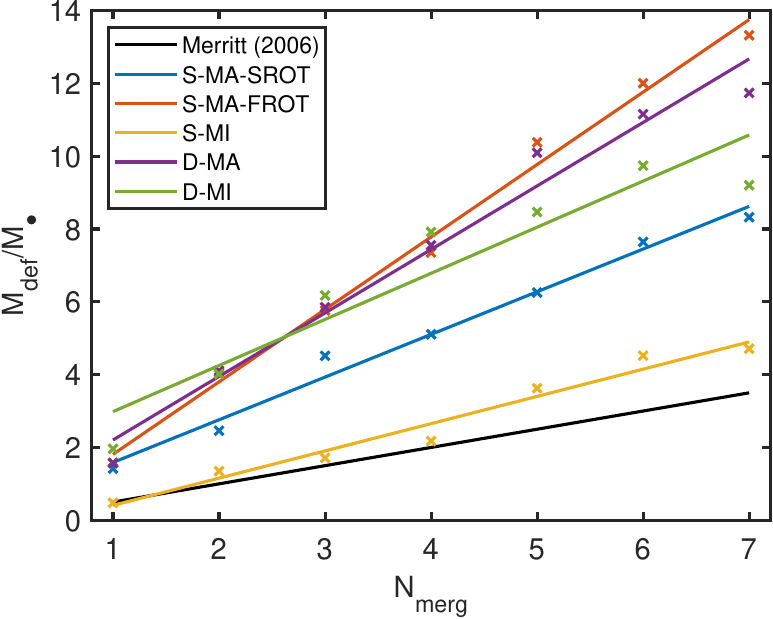}
\caption{The mass deficits $M_\mathrm{def}/M_\bullet$ relative to the SMBH masses $M_\bullet$ after $N_\mathrm{merg}$ merger generations. The deficits are estimated after the SMBH merger has occurred in each simulation.}
\label{fig: mdef}
\end{figure}

\begin{table}
    \centering
    \begin{tabular}{l c c c c}
        \hline
        sequence & $M_\bullet$ & $M_\mathrm{def}$ & $M_\mathrm{def}/M_\star $& $\kappa$\\
         & [$\msol{10^9}$] & [$\msol{10^{10}}$] & & \\
        \hline
        Merritt (2006) & & & & 0.5\\
        S-MA-SROT & $6.8$ & $5.0$ & $0.16$ & 1.17\\
        S-MA-FROT & $6.8$ & $7.1$ & $0.22$ & 1.99\\
        S-MI & $4.8$ & $2.1$ & $0.09$ & 0.75\\
        D-MA & $6.8$ & $7.0$ & $0.22$ & 1.74\\
        D-MI & $4.8$ & $4.2$ & $0.17$ & 1.27\\
        \hline
    \end{tabular}
    \caption{The mass deficits $M_\mathrm{def}$ for the five merger sequences compared to the final SMBH mass $M_\bullet$ and relative to the final stellar mass $M_\star$ after the minor merger generations. The constant $\kappa$ for the mass deficits relative to $M_\bullet$ is defined as $M_\mathrm{def} = \kappa N_\mathrm{merg} M_\bullet$ from Fig. \ref{fig: mdef}. We consistently find higher values for $\kappa$ up to $\kappa\sim2.0$ compared to the value $\kappa=0.5$ of \citet{Merritt2006}.}
    \label{table: merritt}
\end{table}

For the purposes of our analysis, we use a practical definition to determine the mass deficits $M_\mathrm{def}$ of our simulated merger remnants including SMBHs. First, we calculate the influence radius $r_\mathrm{infl}$ of the SMBH using the cumulative stellar mass profile $M_\star(r)$ as $M_\star(r<r_\mathrm{infl}) = M_\bullet$. Next, we define the mass deficit $M_\mathrm{def}$ as the difference of the stellar mass within $r=3\times r_\mathrm{infl}$ from the center of the galaxy in the simulation with and without SMBHs. Thus, our definition of the mass deficit $M_\mathrm{def}$ is 
\begin{equation}
    M_\mathrm{def} = M_\mathrm{\star}^\mathrm{noBH}(r<3\times r_\mathrm{infl}) - M_\mathrm{\star}^\mathrm{BH}(r<3\times r_\mathrm{infl})
\end{equation}
in which $M_\mathrm{\star}^\mathrm{noBH}(r)$ is the cumulative stellar mass profile of a simulated galaxy without SMBHs and $M_\mathrm{\star}^\mathrm{BH}(r)$ is the stellar mass profile of a corresponding simulation including SMBHs.

Work based on numerical simulations of sinking binary SMBHs in stellar bulges has predicted that mass deficit of a simulated core galaxy should correlate with the final SMBH mass $M_\bullet$ and the total number of scouring events the galaxy has experienced \citep{Merritt2006}. The mass deficit $M_\mathrm{def}$ can be thus expressed as
\begin{equation}
    M_\mathrm{def} = \kappa N_\mathrm{merg} M_\bullet.
\end{equation}
\cite{Merritt2006} predicted that this $\kappa$ to be a constant ($\kappa=0.5$) from idealized simulations including a single, spherical stellar bulge with two SMBHs with the mass deficit defined using the difference between the initial and final density profiles.

We calculate the mass deficits from the simulated merger remnants of our five simulation sample pairs S-MA-SROT, S-MA-FROT, S-MI, D-MA and D-MI. The results are displayed in Fig. \ref{fig: mdef} with the slopes $\kappa$ of the individual merger sequence pairs are listed in Table \ref{table: merritt}. The largest mass deficits (compared to $M_\bullet$) occur in samples S-MA-FROT, D-MA and D-MI with $M_\mathrm{def}/M_\mathrm{\bullet}\gtrsim10$ after $N_\mathrm{merg}=7$, as is already evident from the surface density profiles in Fig. \ref{fig: surfacedensity}. The large mass deficits in the initially disky setups D-MA and D-MI originate from the fact that the comparison simulation models without SMBHs acquire high central stellar densities as the cuspy satellites fill the almost flat exponential centers of the models in the mergers. The spherical samples with very eccentric binaries (S-MA-SROT and S-MI) have the smallest mass deficits as more SMBH binary orbital energy ends up being emitted in gravitational wave radiation rather than transforming the central stellar structure of the galaxy. For comparison, the less eccentric SMBH binaries in the spherical sample S-MA-FROT acquire the largest mass deficits of this study as more SMBH binary orbital energy couples to the galaxy rather than escaping via GWs. The mass deficits relative to the stellar masses can also reach relatively high values with high-mass SMBHs and a large number of mergers. In our simulations we find $0.09 \leq M_\mathrm{def}/M_\star \leq 0.22$. Thus, up to approximately $\sim 10$--$20\%$ of the stellar material of the galaxy can be displaced by SMBH core scouring in repeated galaxy mergers.

We find consistently larger values for $\kappa$ than Merritt's value of $\kappa=0.5$, ranging from $\kappa\sim0.75$ (S-MI) to $\kappa\sim2.0$ (S-MA-FROT). We have additionally tested a simplified setup similar to \cite{Merritt2006} and find $\kappa\sim 0.5$ for initially circular binaries in the context of the simple setup. Compared to a multi-component galaxy merger with SMBHs, the setup of \cite{Merritt2006} consists of two SMBHs on a circular orbit at the center of a spherical stellar bulge. Our results suggest that in more realistic merger sequence simulations mass deficits can be up to a factor of $\sim4$ larger compared to the idealized core scouring experiments found in the literature. Similar large mass deficits have also been recently reported by \cite{Partmann2023}.

\subsection{Core-\sersic{} profiles}

We fit the core-\sersic{} profile \citep{Graham2003,Trujillo2004} to the surface density profiles of our simulated galaxies in Fig. \ref{fig: surfacedensity} to quantify the evolution of their central parts. Below the break radius or core size $R_\mathrm{b}$, the core-\sersic{} profile replaces the inner parts of the common \sersic{} profile \citep{Sersic1963,Caon1993} with a typically almost flat central power-law component with a slope of $\gamma$. Above the break radius, the \sersic{} profile remains unchanged. In the terms of surface density $\Sigma(R)$ the core-\sersic{} profile can be written as
\begin{equation}
    \Sigma(R) = \Sigma' \left[ 1 + \left(\frac{R_\mathrm{b}}{R}\right)^\mathrm{\alpha} \right]^\mathrm{\gamma/\alpha} e^{-\mathrm{b} \left[ \left(R + R_\mathrm{b}^\mathrm{\alpha}\right)/R_\mathrm{e}^\mathrm{\alpha} \right]^\mathrm{1/(\alpha n)}}
\end{equation}
in which $n$ is the \sersic{} index and $\Sigma'$ is the overall normalization related to the surface density at the break radius. The parameter $\alpha$ determines the sharpness of the transition between the core and the \sersic{} parts of the profile. The constant $b$ is a somewhat complex function of the core-\sersic{} parameters defined in such a way that $R_\mathrm{e}$ will enclose half of the total mass of the profile \citep{Trujillo2004}.

As in \cite{Rantala2018}, we use the \texttt{cmpfit}\footnote{\texttt{cmpfit}: \href{https://pages.physics.wisc.edu/~craigm/idl/cmpfit.html}{https://pages.physics.wisc.edu/$\sim$craigm/idl/cmpfit.html}} package to perform the core-\sersic{} fits. The radial fitting range for the core-\sersic{} profiles is $0.03$ kpc $<R<30.0$ kpc.

\subsection{Observed vs simulated 2D mass deficits}

\begin{figure}
\includegraphics[width=\columnwidth]{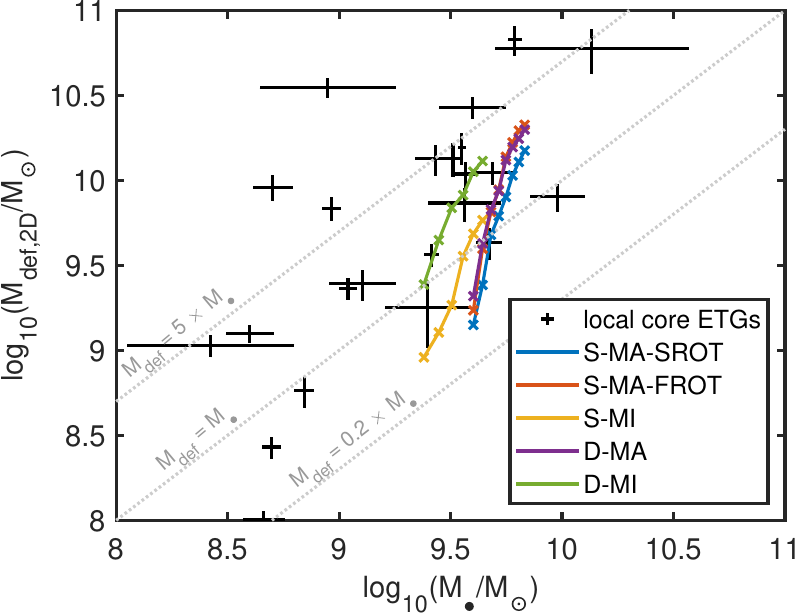}
\caption{The 2D mass deficits $M_\mathrm{def,2D}$ of the simulated merger remnants as a function of the SMBH mass $M_\bullet$ in the merger sequences compared to the mass deficits observed in a sample of local core ETGs \citep{Rusli2013}. The evolution of the 2D mass deficits is steeper than the linear relation $M_\mathrm{def,2D}\propto M_\bullet$.}
\label{fig: coresize-bhmass-obs}
\end{figure}

Observationally, core light deficits are determined by measuring the difference between the actually observed light profile $I(R)$ (or a fit to it) and an inward extrapolation of the \sersic{} profile $I_\mathrm{Sersic}$ that describes the main body of the galaxy \citep{Kormendy2009b,Rusli2013}. These light deficits can be transformed into mass deficits $M_\mathrm{def}$ with an appropriate stellar mass-to-light ratio $\Upsilon=M/L$:
\begin{equation}
    M_\mathrm{def,obs} = 2 \pi \Upsilon \int_\mathrm{0}^\mathrm{\infty} dR R \left[ I_\mathrm{Sersic}(R) -  I(R) \right].
\end{equation}
This \sersic{}-based approach would be desirable for determining the mass deficits of simulated core galaxy models for a consistent comparison with actual observed local ETG mass deficits. However, with increasing number of merger generations the surface density profiles $\Sigma(R)$ in the outer parts approach a power-law, making simple \sersic{} fits with reasonable \sersic{} indices $n<10$ difficult. Hence, a consistent \sersic{}-based mass deficit determination from the simulated galaxies is not straightforward. 

In general, steep power-law outer parts result in unreasonably large (and even diverging) mass deficits when extrapolating the $n>10$ core-\sersic{} fits to the inner parts of the models. In principle, the issue might be avoided by considering non-Hernquist initial models for the outer parts, exploring different merger orbit geometries, or by performing careful multi-component fitting (e.g. section 4 of \citealt{Rusli2013}). We note that also the observed local relic galaxies require multiple \sersic{} components for good surface brightness profile fits \citep{Yildrim2015}. However, here we instead adopt a simple approach for determining the mass deficits from the surface density profiles following section \ref{section: Mdef}. Namely, we define the two-dimensional mass deficit as $M_\mathrm{def,2D}$
\begin{equation}
    M_\mathrm{def,2D} = 2 \pi \int_\mathrm{0}^\mathrm{3 R_\mathrm{b}} dR R \left[ \Sigma_\mathrm{\bullet}(R) -  \Sigma(R) \right]
\end{equation}
in which $\Sigma_\mathrm{\bullet}(R)$ is the scoured model and $\Sigma(R)$ a corresponding simulation profile without SMBHs. We note that this mass deficit determination is not fully consistent with observational mass deficit determination procedures. To estimate mass deficits $M_\mathrm{def,obs}$ of individual galaxies, observers have to rely on a single, scoured core-\sersic{} profile. However, an approach similar to the one presented above has been suggested for a statistical determination of the mass deficits of ETGs as a function of mass \citep{Hopkins2010}. In any case, as shown in Fig. \ref{fig: coresize-bhmass-obs}, the 2D mass deficits $M_\mathrm{def,2D}$ are in the range of observed mass deficits in local ETGs \citep{Rusli2013}. 

The mass deficits $M_\mathrm{def,2D}$ as function of SMBH mass $M_\bullet$ in different simulation samples behave overall in very similar manner and scale steeper than $M_\mathrm{def,2D}\propto M_\bullet$. After the merger sequences the 2D mass deficits are in the range $2.2\lesssim M_\mathrm{def,2D}/M_\bullet\lesssim 3.1$ in the samples S-MA-SROT, S-MA-FROT, D-MA and D-MI, consistent with the mass deficits of the local core ETG population.

Here the prediction of \cite{Merritt2006} based on sinking SMBH binaries in isolated stellar bulges would yield $N_\mathrm{merg}\sim4$--$6$, a somewhat lower number than the true value. The merger sample S-MI that is not scouring a low-density core has $M_\mathrm{def,2D}/M_\bullet\sim1.3$. The good agreement with the simulated 2D and observed mass deficits highlights that observed mass deficits reliably describe the amount of stellar mass displaced from the galactic nuclei due to SMBH binary core scouring even though the true non-scoured profile is observationally unavailable. On the other hand, the good agreement also strengthens the argument for the SMBH binary core scouring origins for the diffuse low-density centers of massive ETGs.

\subsection{Core scaling relations}

\begin{figure*}
\includegraphics[width=\textwidth]{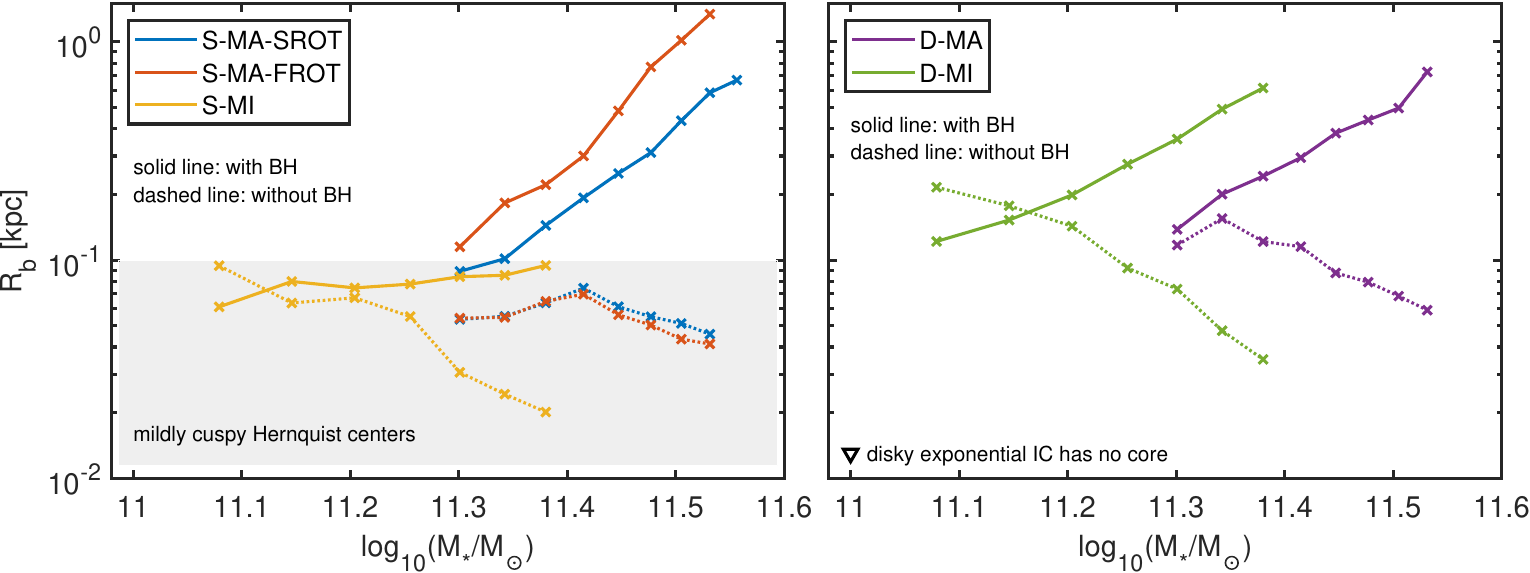}
\caption{The break radii $R_\mathrm{b}$ of the core-\sersic{} fits of the surface density profiles of the simulated merger remnants. The left panel shows the setups with a spherical progenitors (S-MA-SROT, S-MA-FROT, S-MI) while the right panel displays the models with disky progenitors (D-MA, D-MI). The sequences including SMBHs are marked using solid lines while the runs without SMBHs are shown as dotted lines. The shaded region in the left panel indicates the central profiles with Hernquist-like shallow power-law cusps which do not have a distinct flat core despite formally having non-zero break radii $R_\mathrm{b}$ in the core-\sersic{} fits. Only models with SMBHs can grow central flat cores up to sizes of $\sim1$ kpc. The sequence S-MI does not produce a flat core with its very eccentric SMBH binaries as discussed in the main text. Without SMBHs, the almost flat exponential central regions of the initially disky models are filled by stellar material from the in-falling satellites, and their somewhat core-like central initial appearance is lost.}
\label{fig: coresize-stellarmass}
\end{figure*}

\begin{figure}
\includegraphics[width=\columnwidth]{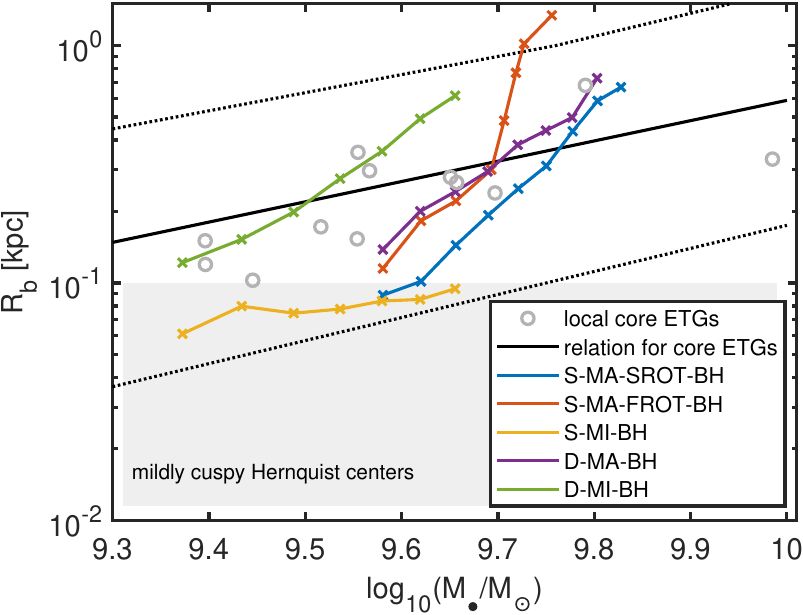}
\caption{The core sizes $R_\mathrm{b}$ as a function of the SMBH mass $M_\bullet$ of the simulated galaxy merger samples compared to the local observed relation of \citet{Thomas2016}. Selected local cored ETGs \citep{Rusli2013,Thomas2016} are shown as open circles. Models with SMBHs (again excluding the model S-MI-BH) can grow large cores up to $R_\mathrm{b}\sim1$ kpc.}
\label{fig: coresize-bhmass}
\end{figure}

\subsubsection{Evolution of the core sizes $R_\mathrm{b}$}

We show the evolution of the core sizes $R_\mathrm{b}$ in the galaxy models in our major and minor merger sequences in Fig. \ref{fig: coresize-stellarmass} as a function of the stellar mass after each merger generation. We first note that the projected Hernquist profile has a shallow power-law center ($\Sigma(R) \propto R^{-\gamma}$ with $\gamma<1$) such that the core-\sersic{} fit obtains a core size $R_\mathrm{b,H} \sim 0.1$ kpc for our isolated initial setups. Hence, we will not consider $R_\mathrm{b}<R_\mathrm{b,H}=0.1$ kpc as actual cores in this study.

For the spherical models without SMBHs (S-MA-SROT, S-MA-FROT, S-MI) the break radius $R_\mathrm{b}$ never exceeds $R_\mathrm{b,H}=0.1$ kpc in the merger sequences. In addition, their $R_\mathrm{b}$ always monotonically decrease starting with the third minor merger generation. Thus, without SMBHs, the initially spherical galaxies will not grow cores in the merger sequences and retain their mildly cuspy central structure.

The disky models D-MA and D-MI originate from the exponential progenitor setup host-D with an exponential surface density profile which by definition does not have a core in the core-\sersic{} sense as there is no break in the profile. After a single major merger without SMBHs, the setup D-MA has a small core with $R_\mathrm{b}\sim0.12$ kpc. Similarly, in the setup D-MI without SMBHs the simulated galaxy has a break radius of $R_\mathrm{b}\sim0.22$ kpc after the first minor merger. However, in the absence of SMBH core scouring in the subsequent mergers, the small cores are filled and $R_\mathrm{b}$ steadily decreases towards the later merger generations. After the entire merger sequence the two models have very small core sizes of $R_\mathrm{b} \sim 0.059$ kpc (D-MA) and $R_\mathrm{b} \sim 0.035$ kpc (D-MI), respectively. Even though the disky progenitor galaxies have almost flat core-like exponential centers, the core-like central feature is erased by violent relaxation in repeated galaxy mergers if no SMBHs are present.

With SMBHs, the model S-MI-BH does not form a core (even though $M_\mathrm{def}>0$) as discussed in the previous sections. The break radius of the model is always close to $R_\mathrm{b}\sim R_\mathrm{b,H}=0.1$ kpc. This is not surprising as the surface density profiles of the model after each merger generation remain practically unchanged, as seen already in Fig. \ref{fig: surfacedensity}.

With SMBHs, the spherical sequences S-MA-SROT-BH and S-MA-FROT-BH including a major merger grow large, diffuse, flat central cores reaching up to kpc in size. After seven minor mergers, the galaxy models have core sizes of $R_\mathrm{b} \sim 0.58$ kpc (S-MA-SROT-BH) and $R_\mathrm{b} \sim 1.34$ kpc (S-MA-FROT-BH), respectively, consistent with massive ETGs with large cores in the local Universe (e.g. \citealt{Lauer2007,Thomas2016}). Reaching the sizes of the largest observed galaxy cores ($R_\mathrm{b}\sim$ a few kpc) potentially requires a subsequent merger of two already cored galaxies \citep{Mehrgan2019,Dullo2019}.

With SMBHs, the disky setups D-MA-BH and D-MI-BH initially have $R_\mathrm{b}\sim R_\mathrm{b,H}$ after their first mergers. In the subsequent mergers including SMBHs, the break radius increases with each merger generation, reaching $R_\mathrm{b} \sim 0.73$ kpc (D-MA-BH) and $R_\mathrm{b} \sim 0.62$ kpc (D-MI-BH) after seven minor mergers. The core sizes are very similar compared to especially the spherical model S-MA-SROT-BH.

\subsubsection{The $M_\bullet$--$R_\mathrm{b}$ relation}

We present the evolution of the simulated galaxies in the $M_\bullet$--$R_\mathrm{b}$ plane in Fig. \ref{fig: coresize-bhmass}. For comparison to the simulation data, we display the core sizes and SMBH masses of a number of observed local ETGs \citep{Rusli2013,Thomas2016}. In addition, we show the observed relation between the core sizes $R_\mathrm{b}$ and the SMBH masses $M_\bullet$ which can be expressed as
\begin{equation}
    \log_\mathrm{10}\left( \frac{M_\bullet}{M_\odot} \right) =  a + b\times\log_\mathrm{10}\left( \frac{R_\mathrm{b}}{\text{kpc}} \right).
\end{equation}
For the coefficients $a$ and $b$ of the relation \cite{Thomas2016} provides values $a=10.27\pm0.51$ and $b=1.17\pm0.14$. 

The observed relation between the core size and the SMBH mass is not especially tight, and all models in which SMBHs scour cores (S-MA-SROT-BH, S-MA-FROT-BH, D-MA-BH, D-MI-BH) are consistent with \cite{Thomas2016}. For the models S-MA-BH the small break radii $R_\mathrm{b}\sim R_\mathrm{b,H}$ are consistent with the mildly cuspy Hernquist centers and are not considered as actual cores. All merger sequences forming cores have their break radii $R_\mathrm{b}$ for a given SMBH mass $M_\bullet$ initially lying below the observed relation. Importantly, in the simulated merger sequences with cores the core size $R_\mathrm{b}$ grows steeper as a function of $M_\bullet$ than the observed relation. Thus, each sequence with a core will eventually cross the relation, and finally be far above it after a large number of minor mergers. In our simulations, the galaxy models reach the observed $M_\bullet$--$R_\mathrm{b}$ relation after $3$--$5$ minor mergers. Only the model S-MA-FROT-BH grows a core large compared to the observed relation, this occurs after $6$--$7$ minor mergers.

We note that the evolution of the simulated galaxies in the $M_\bullet$--$R_\mathrm{b}$ plane depends on the slope of the initial density profile. In general, the progenitors of local massive ETGs are thought to have density profiles steeper than the Hernquist profile up to $\gamma\sim2.0$. The Hernquist model ($\gamma=1.0$) adopted for this study primarily for computational reasons results in more rapid increase of $R_\mathrm{b}$ as a function of $M_\bullet$ compared to initially cuspier profiles such as the $\gamma=3/2$ model as shown by \cite{Rantala2018} in their Fig. 11. The cuspy $\gamma=3/2$ models move closely along the observed local $M_\bullet$--$R_\mathrm{b}$ relation after each merger, indicating no redshift evolution of the core-SMBH relations for such models.

Our results indicate that it is possible for a $z=2$ compact nugget-like ETGs to reach the $z=0$ mass-size relation and simultaneously form a core consistent with the observed $M_\bullet$--$R_\mathrm{b}$ relation after $6$--$7$ minor mergers. The required number of mergers is consistent with empirical galaxy formation models which predict that the minor merger rate hardly exceeds $\sim1$ Gyr$^{-1}$ at $z\lesssim2$ \citep{Oleary2021}. The presence of SMBHs is crucial for the scenario as without SMBHs the galaxy merger remnants do not form cores, even when starting from core-like exponential disky setups.

The slopes and offsets of the correlations between $M_\bullet$ and the mass deficit or the core size probably also contain valuable information about the details of the merging histories of massive galaxies. However, a more quantitative comparison with the simulations would require a careful analysis of potential systematics between observations and simulations and would also require a more thorough exploration of various evolutionary scenarios. This is beyond the scope of this study.

\section{Core scouring and the global structure of ETGs}\label{section: scouring-global}

\subsection{Large mass deficits can increase the effective radii of galaxies}\label{section: scouring-rh}

\begin{figure}
\includegraphics[width=\columnwidth]{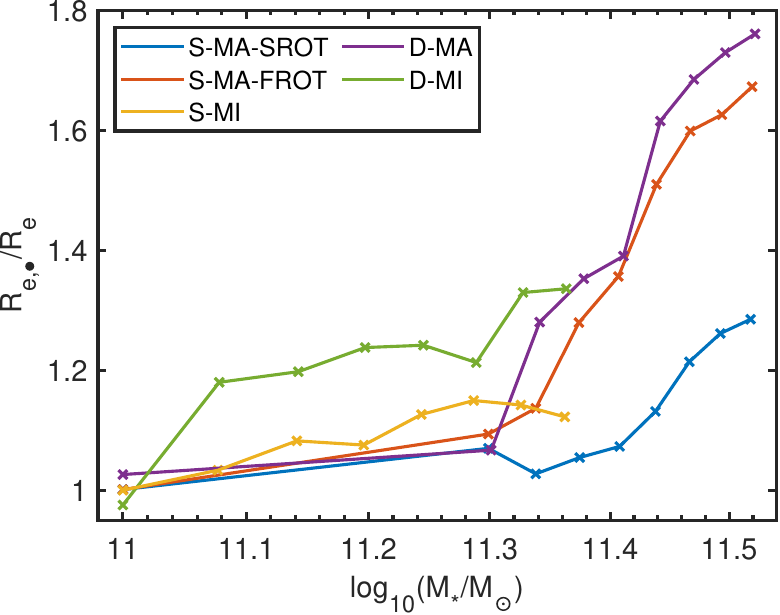}
\caption{The relative increase of the effective radii $R_\mathrm{e,\bullet}/R_\mathrm{e} > 1$ of the galaxies in the merger sequences as a function of increasing stellar mass (merger generation) comparing mergers with and without SMBHs. The additional size growth due to SMBHs is especially prominent in merger samples S-MA-FROT and D-MA, reaching values up to $R_\mathrm{e,\bullet}/R_\mathrm{e} \sim 1.76$ after a single major merger and seven minor mergers.}
\label{fig: delta-rh-bh-simulations}
\end{figure}

In Section \ref{section: reach-mass-size-relation} we pointed out that merger sequences involving SMBHs grow their effective radius faster in the minor merger sequences than their counterparts without SMBHs, i.e. $R_\mathrm{e,\bullet}/R_\mathrm{e} > 1$. In other words, for the size growth relation $R_\mathrm{e} \propto M_\star^\alpha$, the power-law index $\alpha$ is larger in simulations including SMBHs, thus $\alpha_\bullet-\alpha>0$.

In Fig. \ref{fig: delta-rh-bh-simulations} we show the increase of the galaxy size $R_\mathrm{e,\bullet}/R_\mathrm{e}$ in the simulations with SMBHs relative to the simulations without SMBHs as a function of the increasing galaxy stellar mass in the merger simulations. In the three major merger simulations of this study the extra size growth due to SMBHs is mild with $R_\mathrm{e,\bullet}/R_\mathrm{e}\lesssim1.09$. However, in the minor merger sequences the additional size growth can be substantial, especially in the setups S-MA-FROT and D-MA, with the ratio of the effective radii reaching values of up to $R_\mathrm{e,\bullet}/R_\mathrm{e}\sim1.76$. The effect is moderate in the sequences S-MA-SROT ($R_\mathrm{e,\bullet}/R_\mathrm{e}\sim1.29$) and D-MI ($R_\mathrm{e,\bullet}/R_\mathrm{e}\sim1.34$). The sample S-MI which does not form a core with SMBHs even though having $M_\mathrm{def}>0$ has the smallest relative effective radius size increase of $R_\mathrm{e,\bullet}/R_\mathrm{e}\sim1.12$ after seven minor mergers.

\begin{figure*}
\includegraphics[width=0.92\textwidth]{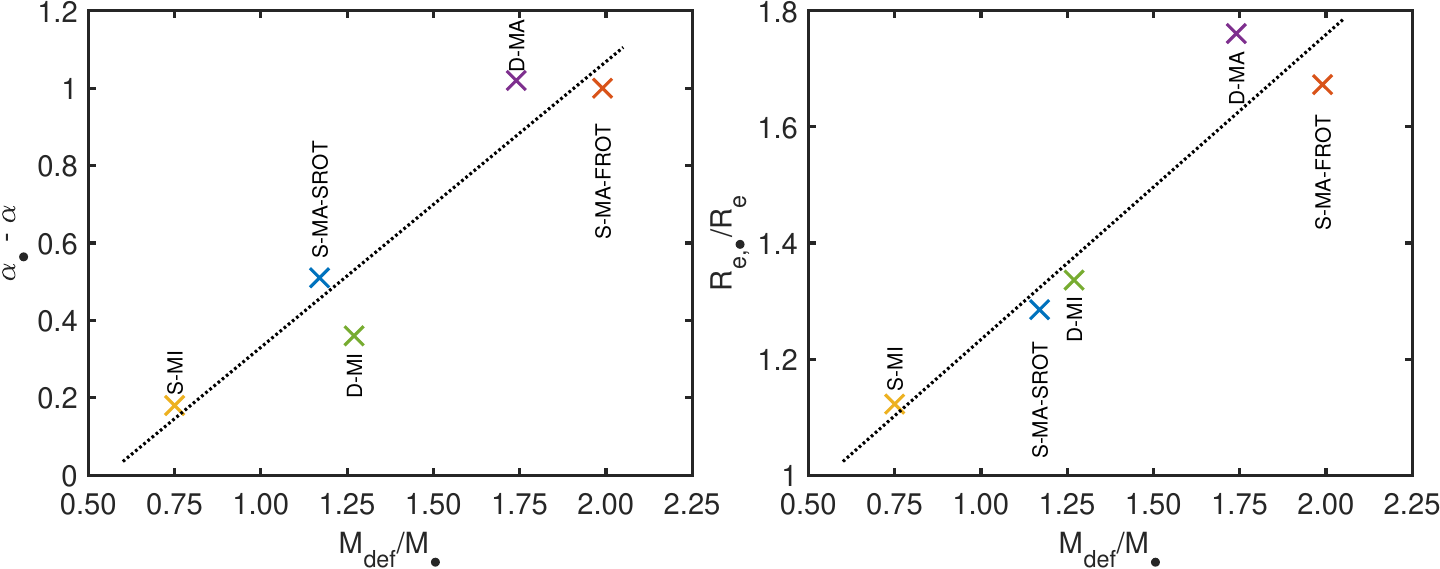}
\caption{The correlation of the mass deficit $M_\mathrm{def}/M_\bullet$ and the additional galaxy size growth as measured by the growth index difference $\alpha_\bullet-\alpha$ (left panel) and the ratio of the effective radii $R_\mathrm{e,\bullet}/R_\mathrm{e}$ (right panel). The dashed lines in each panel show a simple linear fit between the quantities with the fit coefficients are listed in the main text. The displayed results are after seven generations of minor mergers.}
\label{fig: diff-ratio-deficit}
\end{figure*}

Our working hypothesis is that the stellar material displaced by the SMBH core scouring is responsible for the observed size growth effect. We perform a simple test in Appendix \ref{section: appendix-sims-byhand} in which we show that the increase of galaxy size indeed stops if the SMBH binary is merged in the simulation by hand. If the core scouring is not responsible for the observed galaxy size growth in the merger sequences there should be no correlation between the mass deficits $M_\mathrm{def}$ and the size growth diagnostics $R_\mathrm{e,\bullet}/R_\mathrm{e}$ and $\alpha_\bullet-\alpha$. However, we find such correlations between $M_\mathrm{def}$, $R_\mathrm{e,\bullet}/R_\mathrm{e}$ and $\alpha_\bullet-\alpha$. 

We present the size growth - mass deficit correlation results in Fig. \ref{fig: diff-ratio-deficit}. Both $R_\mathrm{e,\bullet}/R_\mathrm{e}$ and $\alpha_\bullet-\alpha$ scale linearly with the final mass deficits of the galaxy models from Table \ref{table: merritt}. The best-fit coefficients for the linear relations with (95\% confidence intervals) are
\begin{equation}
    \alpha_\bullet-\alpha = a + b \times \frac{M_\mathrm{def}}{M_\bullet}
\end{equation}
with $a = 0.71 \pm 0.15$ and $b = 0.53\pm 0.10$ for the size growth index difference and
\begin{equation}
    \frac{R_\mathrm{e,\bullet}}{R_\mathrm{e}} = c + d \times \frac{M_\mathrm{def}}{M_\bullet}
\end{equation}
with $c=-0.41\pm0.21$ and $d=0.74\pm0.14$ for the effective radius ratio. We proceed to explain this linear relationship between the increase of $R_\mathrm{e}$ and the mass deficit $M_\mathrm{def}$ in the next section.

\subsection{A simple analytic model for size evolution due to SMBH scouring}

In the previous section we established the linear dependence of the additional effective radius growth $R_\mathrm{e,\bullet}/R_\mathrm{e}$ in simulations including SMBHs on the mass deficits $M_\mathrm{def}$. We present a simple scouring model to qualitatively explain why the mass deficit can increase the effective radius of a galaxy and why the dependence is linear. The simple scouring model can be summarized with the following points.
\begin{itemize}
    \item A galaxy with a total mass $M$, a cumulative mass profile $M(r)$ and an initial effective radius $R_\mathrm{e}$ corresponding to a half-mass radius $r_\mathrm{h}$ is scoured by one or more SMBH binaries with the final mass deficit of $M_\mathrm{def}$. Stellar mass is removed from the nucleus of the galaxy in such a manner that the central density profile $\rho(r)$ is flat within the core radius.
    \item Case one: scoured stars remain bound. The material scoured from the core is deposited in the outer parts ($r>r_\mathrm{h,old}$) of the galaxy. In this case, a mass of $M_\mathrm{def}\ll M$ is removed from inside the original $r_\mathrm{h,old}$ and added outside it. The new half-mass radius $r_\mathrm{h,new}$ is determined by the criterion $M(r_\mathrm{h,new}) = M(r_\mathrm{h,old}) + M_\mathrm{def}$. Note that in the expression $M(r)$ is the original cumulative mass profile.
    \item Case two: scoured stars get unbound. The scoured material becomes completely unbound of the galaxy, and the mass of the galaxy is decreased by $M_\mathrm{def} \ll M$. The material ejected from within $r_\mathrm{h,old}$ is permanently removed, and the new half-mass radius is calculated as $M(r_\mathrm{h,new}) = M(r_\mathrm{h,old}) + M_\mathrm{def}/2.$
\end{itemize}
We note that our simple model for the effective radius growth due to core scouring does not take into account the scouring of the stellar mass added by galaxy mergers. As such, it is consistent with the \cite{Merritt2006} setup of one stellar bulge with two SMBHs, from which the standard theoretical mass deficits arguments originate. In case of a single merger, one can regard $M(r)$ as the mass profile of the merger remnant before the core scouring.

The two cases can be combined into a single one by introducing a constant $\eta$ which has a value of $\eta=1$ if the scoured material is retained in the outer parts of the galaxy (case one) or $\eta=1/2$ if the material becomes unbound (case two). Denoting the increase of the half-mass radius by $\Delta r_\mathrm{h}=r_\mathrm{h,new}-r_\mathrm{h}$, the relative change of the half-mass radius of the galaxy due to core scouring can be written as 
\begin{equation}\label{eq: rh-increase}
    \frac{\Delta R_\mathrm{e}}{R_\mathrm{e}} = \frac{\Delta r_\mathrm{h}}{r_\mathrm{h}} = \frac{\eta M_\mathrm{def}}{r_\mathrm{h}}  \left[ \left. \derfrac{M(r)}{r} \right|_\mathrm{r=r_\mathrm{h}} \right]^{-1}.
\end{equation}
Most importantly, the relative change depends linearly on the mass deficit $M_\mathrm{def}$, and on the derivative of the cumulative mass profile $M(r)$ at the original half-mass radius $r_\mathrm{h}$. This can be intuitively understood as follows. If the density profile $\rho(r)$ declines sharply near $r_\mathrm{h}$, the mass profile $M(r)$ at $r_\mathrm{h}$ is also very flat. Thus, the half-mass radius must be considerably increased to enclose an additional mass of $\eta M_\mathrm{def}$. The opposite also holds: if the density profile is not very steep at $r_\mathrm{h}$, the mass profile steeply increases there, and even a small increase in the half-mass radius is enough to enclose the extra mass $\eta M_\mathrm{def}$. 

\begin{figure}
\includegraphics[width=\columnwidth]{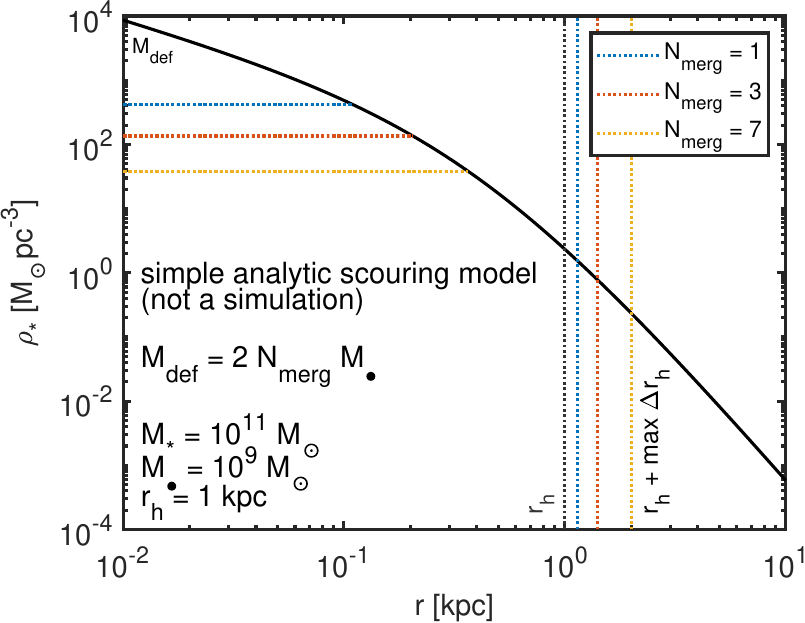}
\caption{A simple scouring model to illustrate the increase of the half-mass radius of a galaxy for large mass deficits. We assume $M_\mathrm{def} = 2 N_\mathrm{merg} M_\bullet$ based on the maximum mass deficits in our simulations listed in Table \ref{table: merritt}. The half-mass radius of the initial model is $r_\mathrm{h}=1.00$ kpc. We assume that the scoured material remains bound and is redistributed in the outer parts of the galaxy, so $\eta=1$ for Eq. \eqref{eq: rh-increase}. After $N_\mathrm{merg}=1$, $N_\mathrm{merg}=3$ and $N_\mathrm{merg}=7$ scouring events, the half-mass radius of the galaxy model has reached $r_\mathrm{h}=1.14$ kpc, $r_\mathrm{h}=1.41$ kpc and $r_\mathrm{h}=2.00$ kpc, respectively.}
\label{fig: scouring-rh}
\end{figure}

For mass profiles that are shallow near $r_\mathrm{h}$, even small mass deficits can result in considerably large relative changes in the half-mass radius due to the derivative term in Eq. \eqref{eq: rh-increase}. We demonstrate this in Fig. \ref{fig: scouring-rh} using the simple scouring model. We assume a galaxy model closely resembling our isolated progenitors: $r_\mathrm{h} = 1$ kpc and $M_\star=\msol{10^{11}}$. We assume that the SMBH of the galaxy with a mass of $M_\bullet=\msol{10^{9}}$ was assembled in $N_\mathrm{merg}=1$--$7$ earlier SMBH mergers, and that the inspirals have removed a mass of $M_\mathrm{def}$ from the center of the galaxy. We calculate the mass deficit using $M_\mathrm{def} = \kappa N_\mathrm{merg} M_\bullet$ with $\kappa=2.0$ motivated by the mass deficits in our simulation sequences listed in Table \ref{table: merritt}. For the fate of the scoured stars we assume that they remain bound in the outer parts of the galaxy model, i.e. $\eta=1$ in Eq. \eqref{eq: rh-increase}. We evaluate the relative effective radius change due to core scouring from \eqref{eq: rh-increase} for the Hernquist profile. This results in a simple relation
\begin{equation}\label{eq: rh-increase-hernquist}
\frac{\Delta R_\mathrm{e}}{R_\mathrm{e}} = \left( 2 + \sqrt{2}\right) \frac{\eta M_\mathrm{def}} {M_\star} \sim 3.41 \times \frac{\eta M_\mathrm{def}} {M_\star}
\end{equation}
with $\Delta R_\mathrm{e}/R_\mathrm{e} \propto M_\mathrm{def}/M_\star$. Inserting the maximum mass deficit $M_\mathrm{def}/M_\star=0.22$ from our simulations (D-MA-BH) in Section \ref{section: Mdef}, we have $\frac{\Delta R_\mathrm{e}}{R_\mathrm{e}} \sim 0.75$ for $\eta=1.0$. This corresponds to a value of $R_\mathrm{e,\bullet}/R_\mathrm{e}\sim1.75$ which is in remarkably good agreement with the actual simulation result $R_\mathrm{e,\bullet}/R_\mathrm{e}\sim1.76$. For the merger sequences S-MA-SROT-BH, S-MA-FROT-BH and S-MI the simple scouring model predicts $R_\mathrm{e,\bullet}/R_\mathrm{e} \sim 1.27$--$1.55$ (simulation result $1.29$, S-MA-SROT-BH), $1.37$--$1.75$ ($1.67$, S-MA-FROT-BH), $1.29$--$1.58$ (1.33, D-MI-BH). The agreement between the simple model and the simulation results is in general very good. The model somewhat overpredicts ($1.15$--$1.31$) the size increase for the sequence S-MI-BH ($1.12$) which did not scour a flat, low-density core.

We emphasize that the spatial galaxy size increase due to mass displacement by core scouring is in general only significant for SMBHs above the $M_\bullet$--$M_\star$ relation, but can reach surprisingly large values for over-massive SMBHs. For SMBHs near the relation ($M_\bullet\sim 1.4\times10^{-3} M_\star$; \citealt{Häring2004}) the relative size increase is
\begin{equation}
    \frac{R_\mathrm{e,\bullet}}{R_\mathrm{e}} \sim 1.0 + 0.01 N_\mathrm{merg} \text{\hspace{3cm}(typical ETGs)}
\end{equation}
assuming the Hernquist profile, $\kappa=2$ for the mass deficit and $\eta=1$ for the scoured material. Thus, for typical ETGs the effect is at most at a few percent level. However, for SMBHs above the $M_\bullet$--$M_\star$ relation the relative increase may reach larger values due to the linear dependence of the size increase on the mass deficit $M_\mathrm{def} \propto M_\bullet$. For SMBHs which lie a factor of $10$ above the $M_\bullet$--$M_\star$ relation we have
\begin{equation}
    \frac{R_\mathrm{e,\bullet}}{R_\mathrm{e}} \sim 1.0 + 0.1 N_\mathrm{merg} \text{\hspace{0.3cm}(SMBH by a factor of 10 over-massive)}
\end{equation}
which can easily lead to galaxy size increases of tens of percents or more. An over-massive SMBH by a factor of $20$ compared to the $M_\bullet$--$M_\star$ relation only requires $N_\mathrm{merg}\sim5$--$6$ past mergers for a size increase by a factor of $2$.

It is expected that the progenitors of present-day massive ETGs had central density profiles steeper $\gamma$ than the Hernquist model. We generalize Eq. \eqref{eq: rh-increase-hernquist} and show the results in Appendix \ref{section: appendix-gamma}. The galaxy size increase due to core scouring is enhanced in more cuspy models. However, the relative size increase depends on the exact value of $\gamma$ only mildly in the range $0 \leq \gamma \lesssim 2.5$. For example, the \cite{Jaffe1983} model ($\gamma=2$) grows in size only $\sim17\%$ more than the Hernquist model ($\gamma=1.0$). Thus, the mass deficit $M_\mathrm{def}$ itself rather than the exact progenitor profile is the main driver of the relative galaxy size increase due to SMBH core scouring.

\section{Shape and rotation properties of ETGs}\label{section: shape-rotation}

\subsection{Galaxy shapes}

\begin{figure}
\includegraphics[width=\columnwidth]{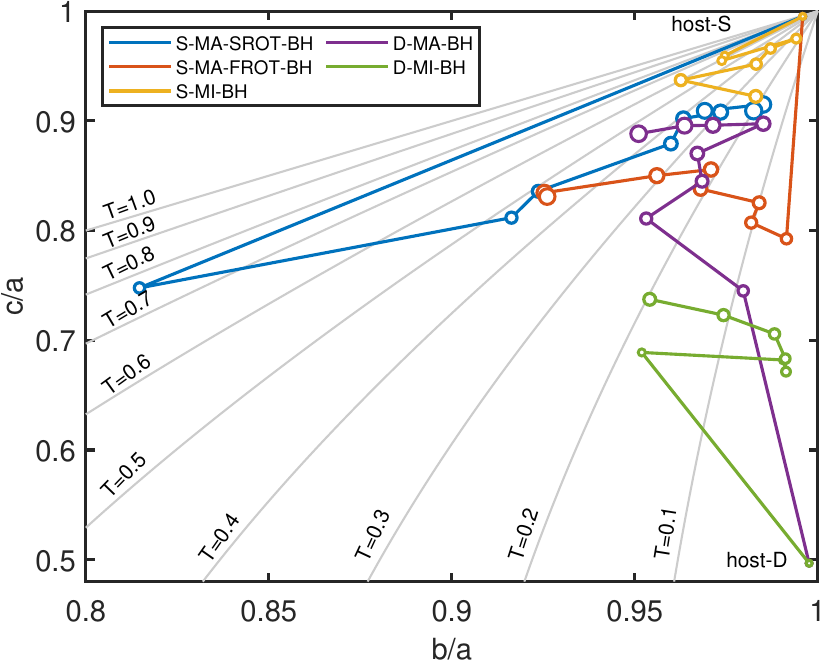}
\caption{The spherical, axisymmetric or triaxial shapes of the initial galaxy models (host-D, host-S) and merger remnants.}
\label{fig: shape_inertiatensor}
\end{figure}

We characterize the three-dimensional shapes of the stellar components of our galaxy models which include SMBHs after each merger generation. The axis ratios $b/a$ and $c/a$ (with $a\geq b \geq c$) within $r_\mathrm{h}$ are determined by calculating the shape (or inertia) tensor of the stellar particle distribution and its eigenvalues (e.g. \citealt{Jesseit2005,Binney2008}) following the recipe of \cite{Zemp2007}. The triaxiality parameter $T$ can in turn be calculated from the axis ratios $b/a$ and $c/a$ as
\begin{equation}
    T = \frac{1-\left( b/a\right)^2}{1-\left( c/a\right)^2}.
\end{equation}
We show the shapes of our galaxy models after each merger generation in the $b/a$--$c/a$ plane in Fig. \ref{fig: shape_inertiatensor}. By construction, the spherical galaxy progenitor host-S has $b/a = c/a = 1$ and for the disky progenitor $b/a=1$ and $c/a=0.5$.

The model S-MA-SROT-BH has a shape of $b/a=0.81$, $c/a=0.75$ ($T=0.77$) after its major merger. This merger remnant is the most triaxial galaxy of the study. In the minor mergers, both $b/a$ and $c/a$ steadily increase for the model up to $b/a=0.98$, $c/a=0.91$ ($T=0.20$). On the contrary, the setup S-MA-FROT-BH is axisymmetric ($b/a=0.99$, $c/a=0.79$) rather than triaxial after its major merger. In the subsequent minor mergers, the model becomes somewhat more triaxial as $b/a$ decreases to a value of $b/a=0.93$ while $c/a$ changes even less with a final value of $c/a=0.83$ ($T=0.46$). The merger remnants in the merger sequence S-MI-BH with spherical progenitors and without a major merger are the most spherical galaxies of the simulations of this study with $c/a>0.92$ for each remnant.

The models originating from disky progenitors (D-MA-BH, D-MI-BH) are all axisymmetric as $b/a>0.95$ for all the merger remnants. The model D-MA-BH has $c/a=0.75$ after the major merger and $c/a=0.89$ at the end of the minor merger sequence. The setup D-MI-BH remains the most disk-like among the merger remnants of the study with its axisymmetric shape with $c/a=0.74$ after the minor mergers.

\subsection{SMBH binary core scouring as a global process: the orbit structure of scoured ETGs}

\begin{figure}
\includegraphics[width=\columnwidth]{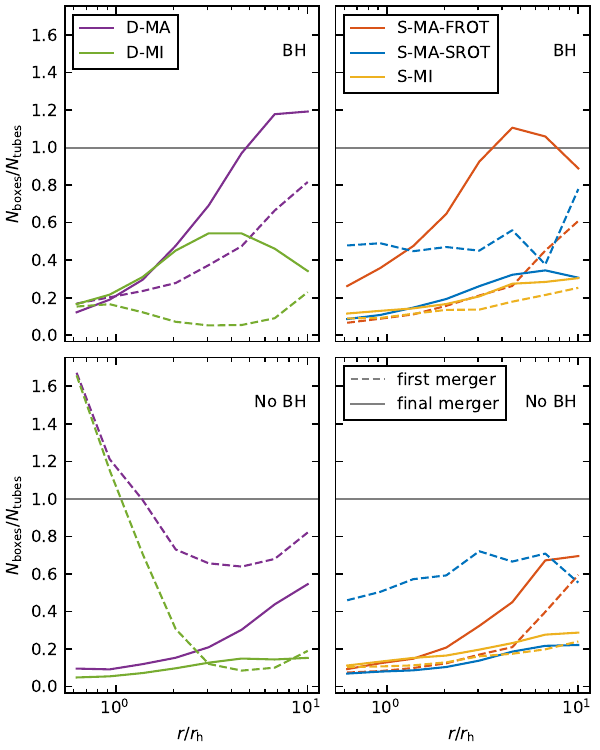}
\caption{The ratio of box orbits to tube orbits $\boxtube$ as a function of radius, scaled to the half-mass radius of each galaxy.
The top row shows $\boxtube$ for disk progenitors (left) and spherical progenitors (right), both with a BH, and the bottom row shows $\boxtube$ for disk progenitors (left) and spherical progenitors (right), both without a BH.
In all panels, the dashed line represents $\boxtube$ following the first merger of the sequence, and the solid line $\boxtube$ following the final merger of the sequence.
}
\label{fig:ratio_bt}
\end{figure}

SMBH binaries preferentially eject stars on close-to-radial orbits with little angular momentum. Stars with low angular momentum orbits do not spend all their time within the central region, but instead spend substantial periods of time further out in the galaxy. Thus, SMBH core scouring removing these orbits can be in a sense viewed as a process globally changing the orbit structure of the galaxy, not just altering the core within $R_\mathrm{b}$ (see \citealt{Frigo2021}). In the previous section we demonstrated that core scouring by overmassive SMBHs can alter the global properties of the host galaxies, such as their effective radii $R_\mathrm{e}$. Next, we study this process in more detail using orbit analysis techniques.

We follow the methodology presented in \citet{Frigo2021} to construct an analytical potential in which we can integrate the stellar particles to classify the particles into one of six orbital families.
To determine which family a particular stellar orbit belongs to, the frequency of the particle's motion is determined along each principal axis of the galaxy's inertia tensor through a Fourier transform. 
Briefly, these orbit families are: 
\begin{enumerate}
    \item $\pi$-box orbits, which have no net angular momentum and no resonant motion along the three principal axes,
    \item boxlet orbits, which also have no net angular momentum, but do show resonant motion along the three principal axes,
    \item $x$-tube orbits, which rotate about the major axis of the galaxy,
    \item $z$-tube orbits, which rotate about the minor axis of the galaxy,
    \item rosette orbits, which are typically associated with a spherically-symmetric potential and conserve each component of the angular momentum vector, and
    \item irregular orbits, where the particle is not bound by any integrals of motion.
\end{enumerate}
Having classified the stellar particles into the above orbital families, we group the families into two broad classes: orbits without angular momentum ($\pi$-box and boxlet orbits) which we term `boxes', and orbits with some degree of angular momentum (rosette, $x$-, and $z$-tube orbits) which we term `tubes'.

Particles are then binned radially into eight logarithmically-spaced bins spanning $0.5r_\mathrm{h}$ to $12r_\mathrm{h}$, and we determine the ratio of box orbits to tube orbits $\boxtube$ for each radial bin.
This analysis is done for the first snapshot after the first merger (this may be either a major or minor merger depending on the merger sequence) and the first snapshot after the final merger (which by construction is always a minor merger).
We show the box-tube ratio in Fig. \ref{fig:ratio_bt}, with dashed lines indicating the ratio following the first merger, and solid lines indicating the ratio following the final merger.
A general trend we observe is that the simulated merger remnants tend to be significantly dominated by tube orbits at small radii compared to box orbits following the final merger.
We discuss each simulated merger remnant in detail below.

Looking first at disky mergers with a BH, we see that for $r\leq r_\mathrm{h}$, little change occurs for the ratio of box to tube orbits. 
This is in agreement with the galaxy shapes in Fig. \ref{fig: shape_inertiatensor}, where for both the D-MA-BH and D-MI-BH sequences the $b/a$ fraction is consistent  after the first merger and after the final merger. 
For radii $r>r_\mathrm{h}$, both D-MI-BH and D-MA-BH increase the fraction of box orbits during their merger sequence.
The increase in $\boxtube$ beyond $r_\mathrm{h}$ can naturally be understood as the stellar particles that have interacted with the SMBH being deposited at larger radii, as described in the analytical scouring model in section \ref{section: scouring-global}. 
In particular, many tube orbits that interact with a BH binary are converted to either radial box orbits, or unbound from the system completely \citep{Frigo2021}.
In both cases, $\boxtube$ of the final merger remnant is raised compared to the ratio following the first merger. 

Turning our attention to the disky mergers without a BH, we see a distinct difference to the case with a BH. 
Following the first merger, both D-MA and D-MI are dominated by box orbits within $r_\mathrm{h}$, with $\boxtube$ in excess of unity.
Following the final merger, the fraction of box orbits within $r_\mathrm{h}$ has decreased to less than 0.1: more than an order of magnitude compared to the first merger.
In both instances, we find that the central stellar densities of the systems are very high. Comparing the snapshot following the final merger in the D-MA case with the equivalent D-MA-BH case, we see that the there is an equal amount of mass within $\sim10^{-2}\:\mathrm{kpc}$ between the two simulations.
In the D-MA-BH case, the majority of this mass is contributed by the SMBH, whereas in the D-MA case the mass is due almost exclusively to stellar particles.
Additionally, in the D-MA case, there is more mass than in the D-MA-BH case for radii $10^{-2}\:\mathrm{kpc} \lesssim r \lesssim 10\:\mathrm{kpc}$ due to violent relaxation.
The high central stellar densities provides the necessary gravitational potential to sustain a large fraction of tube orbits, thus lowering the overall value of $\boxtube$ for the D-MA and D-MI systems following the final merger in their respective sequences.

We now consider the spherical progenitor mergers with SMBHs in the top right panel of Fig. \ref{fig:ratio_bt}.
Immediately apparent is that the S-MI-BH merger sequence displays only minor variation in $\boxtube$ following the first and final mergers, taking a value of $\sim 0.1$ for $r\leq r_\mathrm{h}$, and increasingly mildly with radius to $\sim 0.2$ at the outermost radii.
There is a slight increase in $\boxtube$ following the first merger as compared to the final merger.
The case is drastically different for S-MA-FROT-BH and S-MA-SROT-BH merger sequences.
S-MA-FROT-BH displays an evolution in $\boxtube$ between mergers consistent with the disky mergers, where the ratio increases by the final merger compared to the first merger.
By $4r_\mathrm{h}$, $\boxtube$ has exceeded unity, indicating that at large radii box orbits are the dominant orbital family.
Conversely, for the S-MA-SROT-BH sequence, $\boxtube$ decreases from the first merger to the final merger.
Tube orbits always dominate over box orbits at all radii.
The peculiar change in ratio between the first and final mergers is reflected in the galaxy shape plot of Fig. \ref{fig: shape_inertiatensor}.
For D-MA-BH, D-MI-BH, and S-MA-FROT-BH the triaxiality parameter $T$ increases from the first merger to the final merger, with a final value $0.1 \lesssim T \lesssim 0.4$.
For the S-MA-SROT-BH case, the galaxy following the first merger is relatively prolate ($T\sim 0.7$), and becomes increasingly oblate, with a final $T$ of $\sim 0.2$.
Tube orbits, in particular $z$-tube orbits with their regular rotation about the minor axis, are responsible for a more oblate remnant.

We see similar trends for the spherical mergers without SMBHs as for their SMBH merger counterparts, with the exception of the S-MA-FROT box-tube ratio never exceeding unity: for all spherical mergers without a BH, tube orbits dominate over box orbits at all radii.
The magnitude of $\boxtube$ for radii within $r_\mathrm{h}$ is also very similar for the spherical mergers without a SMBH as for the mergers with a SMBH, with the exception of S-MA-FROT.
In the S-MA-FROT-BH sequence, the magnitude of the box-tube ratio is $\sim30\%$, compared to $\sim10\%$ in the S-MA-FROT case.

\subsection{Velocity anisotropy profiles}

\begin{figure}
\includegraphics[width=\columnwidth]{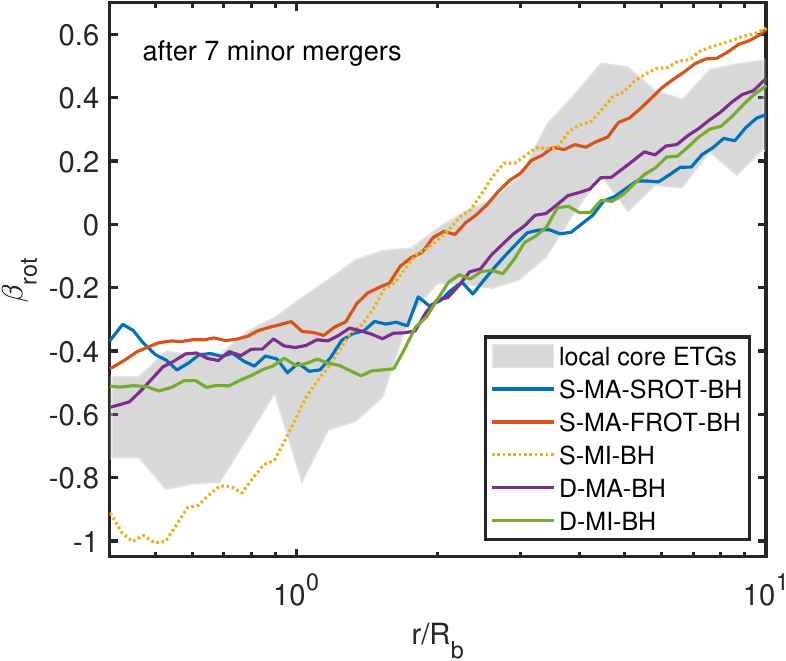}
\caption{The orbital anisotropy parameter $\beta_\mathrm{rot}$ for the simulated merger remnants after seven minor merger generations. The radial coordinate $r$ on the x-axis is scaled with the two-dimensional break radii $R_\mathrm{b}$ of the models, except for the sample S-MI-BH (dashed line) which did not scour a flat central core. For the S-MI-BH sample we use here $R_\mathrm{b}=1$ kpc for visualization purposes. The simulated merger remnants are tangentially biased ($\beta_\mathrm{rot}<0$) within $r\lesssim2 R_\mathrm{b}$. The observed orbital anisotropies of cored, local ETGs \citep{Thomas2014} in the SINFONI BH survey \citep{Saglia2016} are shown as the shaded region.}
\label{fig: beta}
\end{figure}

Next we study the velocity anisotropy structure of our merger remnants after the minor merger generations. The orbital anisotropy parameter $\beta_\mathrm{rot}$ \citep{Thomas2014} we use is defined as
\begin{equation}
\begin{split}
    \beta_\mathrm{rot} &= 1 - \frac{\sigma_\mathrm{rot}^2}{2 \sigma_\mathrm{r}^2}\\
    \sigma_\mathrm{rot}^2 &= \sigma_\mathrm{\theta}^2 + \sigma_\mathrm{\phi}^2 + v_\mathrm{\phi}^2
\end{split}
\end{equation}
using the $\phi$ component of the rotation velocity $v_\mathrm{\phi}$ and the three components of the velocity dispersion $\sigma_\mathrm{r}$, $\sigma_\mathrm{\theta}$ and $\sigma_\mathrm{\phi}$ in spherical coordinates.

We show the orbital anisotropies $\beta_\mathrm{rot}$ of the final merger remnants after seven generations of minor mergers including SMBHs in Fig. \ref{fig: beta}. All the merger remnants are tangentially biased ($\beta_\mathrm{rot}<0$) at separations $r\lesssim2$--$3 R_\mathrm{b}$ from the center due to the prevalence of tube orbits at these radii. Perhaps surprisingly, the model S-MI-BH which did not scour a proper flat core has the lowest central anisotropy, $\beta_\mathrm{rot}\sim-1$. The models S-MA-SROT-BH, S-MA-FROT-BH, D-MA and D-MI have similar $\beta_\mathrm{rot}(r)$ profiles and central values of $-0.6\lesssim\beta_\mathrm{rot}\lesssim-0.3$. We compare these results to galaxies from the SINFONI black hole survey \citep{Saglia2016}. The orbital anisotropy profiles of observed galaxies \citep{Thomas2014} are very consistent with the profiles of the simulated merger remnants.

\subsection{Stellar kinematics in two dimensions}

\begin{figure*}
\includegraphics[width=0.675\textwidth]{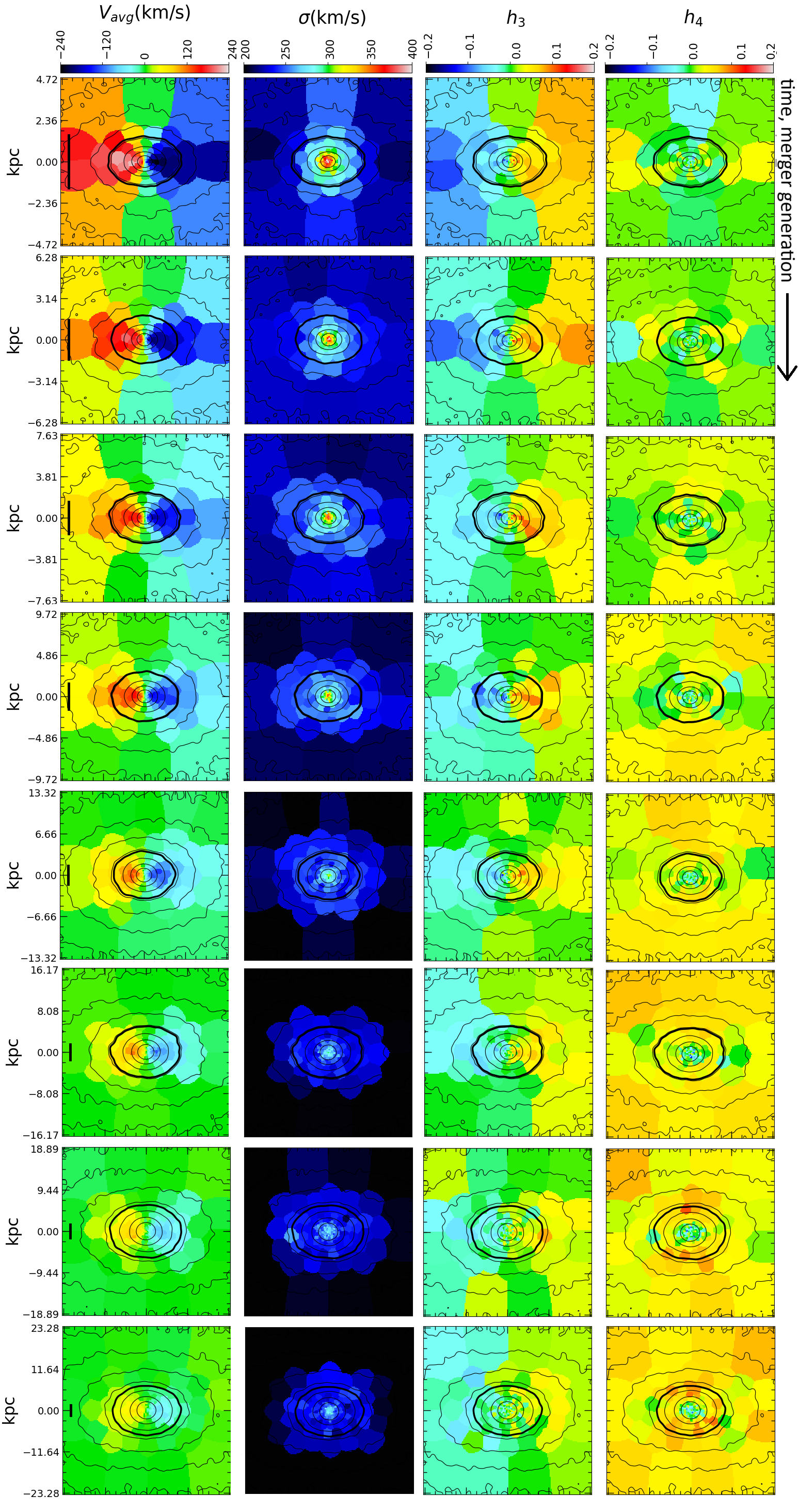}
\caption{Two-dimensional kinematic maps of the simulation sequence S-MA-FROT-BH. The isophote coinciding with the effective radius $R_\mathrm{e}$ of each merger remnant is highlighted. The vertical bars in the left panels mark the extent of $3$ kpc. The spatial extent of each panel is six times the effective radius $R_\mathrm{e}$ of the merger remnant. The various features of the maps are discussed in the main text.}
\label{fig: velmaps-S-MA-FROT-BH}
\end{figure*}

\begin{figure*}
\includegraphics[width=0.675\textwidth]{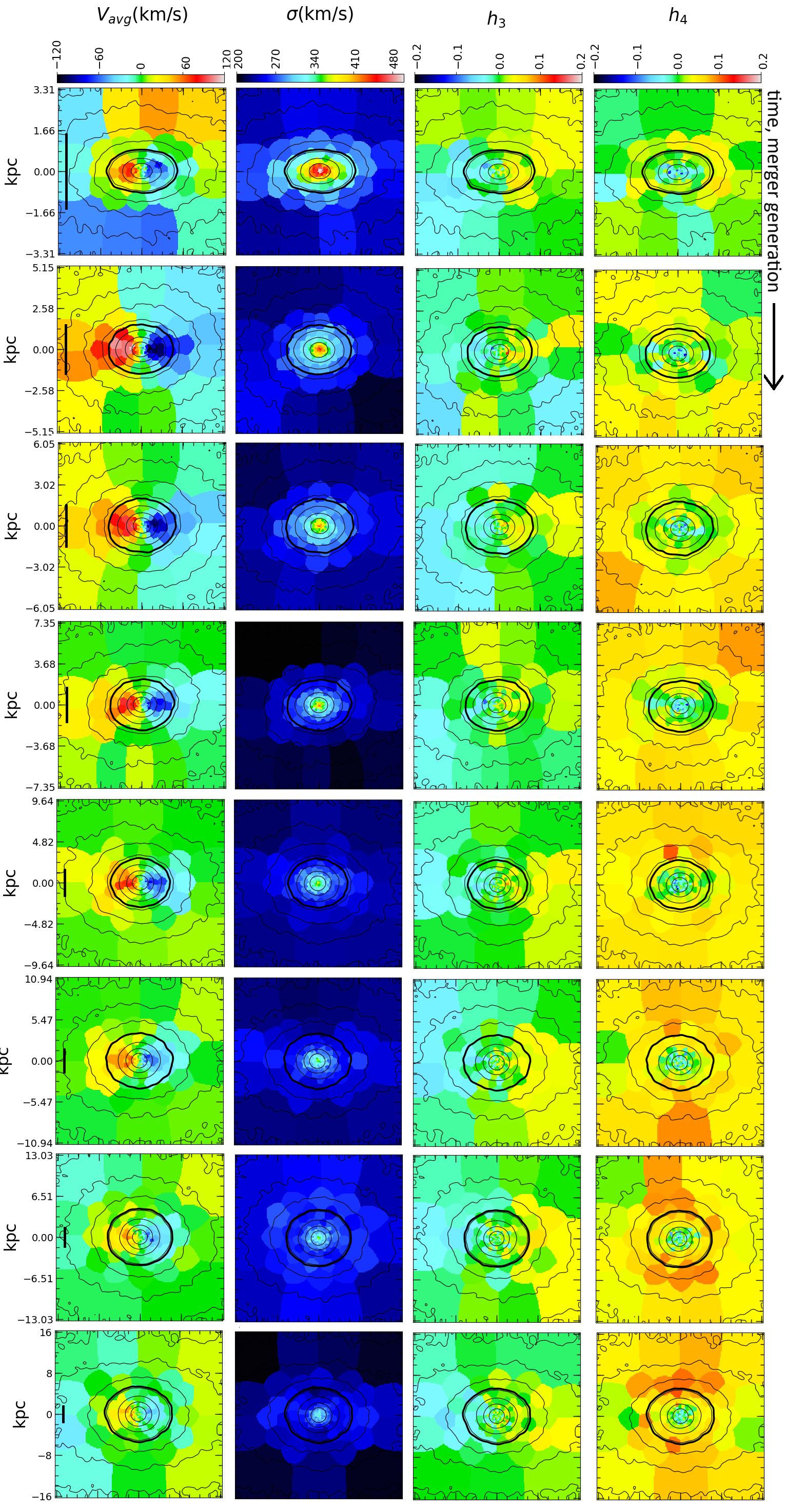}
\caption{Two-dimensional kinematic maps of the simulation sequence D-MA-BH.}
\label{fig: velmaps-D-MA-BH}
\end{figure*}

We show the two-dimensional kinematic maps produced using a modified version of the \texttt{pygad} \citep{Rottgers2020} analysis package of the simulation sequences S-MA-FROT-BH and D-MA-BH in Fig. \ref{fig: velmaps-S-MA-FROT-BH} and in Fig. \ref{fig: velmaps-D-MA-BH}. The two simulation samples were chosen for the figures as they have the strongest rotation signatures among all the runs of this study. The maps are constructed to mimic how the simulated merger remnants would be seen if observed with an integral field unit spectrograph (see e.g. \citealt{Naab2014}). The simulation particle velocity data is first binned on a regular grid with a spatial resolution of $0.03\times R_\mathrm{e}$ of each galaxy. The regular grid is then Voronoi binned into spaxels with a constant signal-to-noise ratio, i.e. the number of particles using a Voronoi tessellation algorithm \citep{Cappellari2003}. Specifically, we use a version of the Voronoi algorithm in a modified version of the \texttt{pygad} package \citep{Rottgers2020}. 

The four panels of Fig. \ref{fig: velmaps-S-MA-FROT-BH} and Fig. \ref{fig: velmaps-D-MA-BH} show in each row, spaxel by spaxel, the line-of-sight velocity $V_\mathrm{avg}$, the velocity dispersion $\sigma$ and the parameters $h_\mathrm{3}$ and $h_\mathrm{4}$ that measure the skewness and kurtosis of the velocity distribution in the spaxels \citep{Gerhard1993,vanderMarel1993}. In general, the anti-correlation of $V_\mathrm{avg}$ and $h_\mathrm{3}$ points towards rotating, disky structures while negative $h_\mathrm{4}$ is a proxy for a tangentially biased velocity distribution with preferentially circular orbits. Finally, the panels also show the two-dimensional flux (surface density) contours with separations of one magnitude.

The kinematic maps of the merger remnants in the simulation sequence S-MA-FROT-BH are shown in Fig. \ref{fig: velmaps-S-MA-FROT-BH}. After the major merger, the remnant is rapidly rotating with $|V_\mathrm{avg}|\sim240$ km/s. The region of fast rotation extends beyond $R=2 R_\mathrm{e}$. The peak $V_\mathrm{avg}$ decreases with increasing number of minor mergers while the region of fast rotation shrinks inside the increasing $R_\mathrm{e}$. After the minor merger generations $|V_\mathrm{avg}|\lesssim70$--$80$ km/s as the mergers from isotropic directions bring in more mass but on average no net angular momentum. The velocity dispersion maps show a single, central peak with $\sigma\sim400$ km/s after the major merger. As the central stellar density is lowered due to core scouring, the central velocity dispersion also decreases, reaching $\sigma\sim275$ km/s after the minor mergers. The $h_\mathrm{3}$ maps anticorrelate with the corresponding $V_\mathrm{avg}$ maps. With increasing number of minor mergers, the $h_\mathrm{4}$ is increasingly more negative in the core regions and increasingly more positive outside it. Finally, the isophotal shapes are in general ellipsoidal (but circular at the center) and turn somewhat more boxy in the minor merger sequences. Overall, the evolution of the isophotal shapes is relatively minor.

After the disk-disk major merger, the merger remnant D-MA-BH has a maximum $|V_\mathrm{avg}| \sim 120$ km/s peaking at $R\sim0.5 R_\mathrm{e}$. Outside the effective radius there is less rotation. In the minor merger generations the rotation weakens, reaching $|V_\mathrm{avg}|\lesssim50$ km/s after seven minor mergers. Similarly, the central velocity dispersion decreases considerably from $\sigma\sim500$ km/s to $\sigma\sim300$ km/s as the initially exponential model transforms into a cored ETG. Again, the $h_\mathrm{3}$ and $V_\mathrm{avg}$ anticorrelate, and $h_\mathrm{4}$ is increasingly negative within the core and positive outside it. The negative central $h_\mathrm{4}$ appears earlier in the minor merger generations compared to the sequence S-MA-FROT-BH. The isophotal shapes are initially ellipsoidal, also near the center, and become increasingly round after each minor merger generation.

We note that our simulated merger remnants lack kinematic misalignments observed in a subset of local, massive ETGs \citep{Franx1991,Krajnovic2011}. This is most probably due to the fact that the satellite galaxies infall from isotropic directions in the merger sequences, preventing the formation of distinct kinematic features.
In general, however, the presence and the structure of kinematic misalignements in the centers of massive core galaxies potentially contain very important clues about the importance of SMBHs for the evolution of massive galaxies \citep{Rantala2019,Neureiter2023}, in particular about the relative importance of major versus minor mergers.

\subsection{The slow and fast rotating ETGs in the merger sequences}

\begin{figure}
\includegraphics[width=\columnwidth]{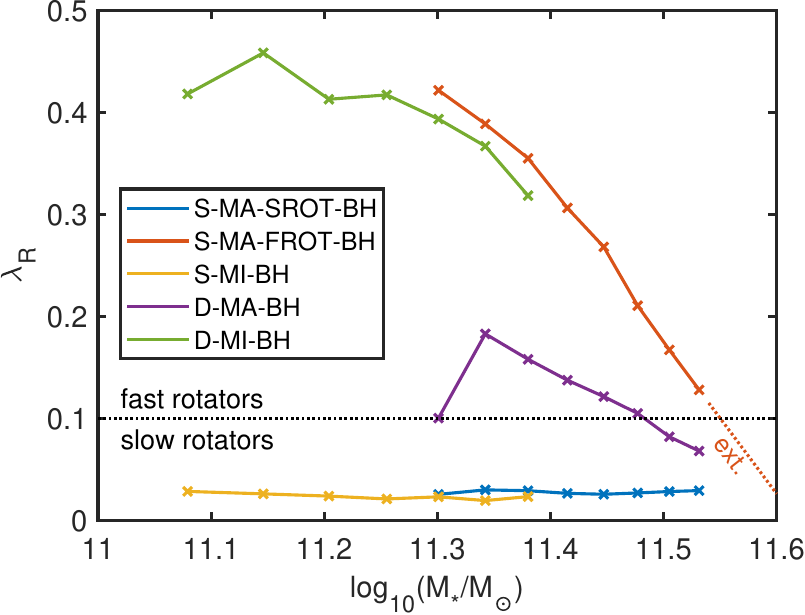}
\caption{The spin parameter $\lambda_\mathrm{R}$ and the classification of our simulated merger remnants into slow ($\lambda_\mathrm{R}<0.1$) and fast ($\lambda_\mathrm{R}>0.1$) rotators. The remnants in the sequences S-MA-SROT-BH and S-MI-BH are always slowly rotating. In the sequences S-MA-FROT-BH and D-MA-BH, the remnants are initially fast-rotating, but become (or will soon become) slow rotators in the minor merger sequences. The remnants in the sequence D-MI-BH remain fast-rotating even after seven minor mergers.}
\label{fig: rotation-lambda}
\end{figure}

We use the spin parameter $\lambda_\mathrm{R}$ as determined from the two-dimensional kinematic data \citep{Rottgers2020} to characterize the rotation properties of our simulated merger remnants. The common classification of ETGs into slow and fast rotators is based on $\lambda_\mathrm{R}$ \citep{Emsellem2007}. Slowly rotating ETGs have $\lambda_\mathrm{R}<0.1$ while for the fast rotators $\lambda_\mathrm{R}>0.1$. Following \cite{Emsellem2007}, the parameter $\lambda_\mathrm{R}$ is defined from the two-dimensional kinematic data as
\begin{equation}
    \lambda_\mathrm{R} = \frac{ \Sigma_\mathrm{i} F_\mathrm{i} R_\mathrm{i} |V_\mathrm{i}| }{ \Sigma_\mathrm{i} F_\mathrm{i} R_\mathrm{i} \sqrt{ V_\mathrm{i}^2 + \sigma_\mathrm{i}^2 }  }
\end{equation}
in which $F_\mathrm{i}$ is the flux (surface density) of each spaxel, $R_\mathrm{i}$ their distance from the center of the galaxy and $V_\mathrm{i}$ and $\sigma_\mathrm{i}$ the line-of-sight velocity and velocity dispersion in each spaxel. The sum runs over all spaxels within the effective radius $R_\mathrm{e}$ of the galaxy. For rotation-dominated systems $\lambda_\mathrm{R}\sim1$ and for non-rotating systems $\lambda_\mathrm{R}\sim0$. Our isolated initial galaxy models have rotation parameters of $\lambda_\mathrm{R}\sim0.0$ (host-S) and $\lambda_\mathrm{R}\sim0.7$ (host-D).

We show the evolution of the rotation parameter $\lambda_\mathrm{R}$ in the simulated merger sequences in Fig. \ref{fig: rotation-lambda}. The galaxy models in the merger sequences S-MA-SROT-BH and S-MI-BH are always slowly rotating with $\lambda_\mathrm{R}<0.04$ for each merger remnant. The disky-disky major merger remnant of the sequence D-MA-BH is marginally fast rotating and increases its rotation parameter to $\lambda_\mathrm{R}\sim0.18$ in its first minor merger. Afterwards, the minor mergers decrease the $\lambda_\mathrm{R}$ bringing in stellar mass but no net angular momentum. After $\sim6$ minor mergers, the remnant D-MA-BH would be classified as a slow rotator. The major merger of the sequence S-MA-FROT-BH results in a fast-rotating remnant due to its merger orbit with $\lambda_\mathrm{R} \sim 0.42$. Each merger generation gradually lowers the $\lambda_\mathrm{R}$ and after seven minor mergers $\lambda_\mathrm{R}\sim 0.13$, a marginal fast rotator. Extrapolating the evolution of $\lambda_\mathrm{R}$ in the minor mergers, we argue that one more minor merger would be required to render the remnant of the sequence S-MA-FROT-BH a slow rotator. Finally, the galaxy models of the sequence D-MI-BH remain fast rotating with $\lambda_\mathrm{R}\sim0.34$ after the minor merger generations.

Most importantly, we have shown that besides the $M_\star$--$R_\mathrm{e}$ size growth and the formation of flat central cores up to $R_\mathrm{b}\sim1$ kpc size, the end products of merger sequences starting from compact early-type galaxies can resemble $z=0$ massive ETGs also in their rotation properties. Both initially rapidly rotating compact ETG models and spherical models gaining rapid rotation due to merger orbit geometry can get rid of their rotation in $6$--$9$ minor mergers. In our minor merger sequences this is due to the fact that on average the minor mergers bring in more stellar mass but no net angular momentum.

\section{Summary and conclusions}\label{section: conclusions}

We have studied the evolution of $z=2$ massive compact early-type galaxies into present-day massive, slowly rotating, cored ETGs in simulated sequences of major and minor galaxy mergers with and without supermassive black holes using the \ketju{} code. Our main result is that $z=2$ compact ETGs (both spherical and disky) can evolve through mergers into $z=0$ galaxies consistent with massive ETGs in their effective radius, core size and rotation properties. The presence of SMBHs in the galaxies is crucial for the models as without SMBHs no low-density core is scoured. We further summarize our findings in detail below.

Compact $z=2$ ETGs can reach the $z=0$ mass-size relation after $6$-$9$ generations minor mergers. Minor mergers are more efficient than major merger in promoting the ETG size growth as suggested by virial arguments \citep{Bezanson2009,Naab2009}. Without SMBHs, we find $1.8\lesssim\alpha \lesssim 2.33$ for the mass growth relation $R_\mathrm{e} \propto M_\star^\alpha$, consistent with the simulations of \cite{Hilz2012,Hilz2013}, with $\alpha\gtrsim2.0$ required to explain the observed ETG size growth \citep{vanDokkum2010} since $z=2$. We find consistently steeper galaxy size growth in simulations including SMBHs with $2.31\lesssim\alpha_\bullet\lesssim 3.10$. Our results highlight the paramount importance of multiple minor mergers including SMBHs on the evolution of massive ETGs.

Repeated minor mergers with SMBHs can scour kpc-size cores into initially compact galaxies, especially if the sequence begins with a major merger. The break radii reached after seven generation of minor mergers are $0.62$ kpc $\lesssim R_\mathrm{b}\lesssim1.34$ kpc and are consistent with local, core ETGs of similar SMBH mass \citep{Rusli2013,Thomas2016}. The model S-MI-BH from a spherical progenitor without a major merger does not scour a core even though its center is less dense than in the corresponding comparison simulations without SMBHs. Hence major mergers are important to explain the entire properties of massive ETGs (see also \citealt{Kluge2023}). The eccentric, unequal-mass binaries of the merger sequence S-MI-BH lose their orbital energy mainly to GW emission and not to the stellar component, and merge without significantly affecting their host nuclei. Without SMBHs, the central stellar densities of the galaxies increase, and the core-like almost flat exponential central parts of the initially disky models are turned into mild power-law cusps. Thus, SMBHs are crucial for the $z=2$ compact ETGs to evolve into the most massive ETGs in the local Universe: without SMBHs, there is no flat central core, even when starting from a core-like exponential progenitor galaxy.

We find 3D mass deficits up to $M_\mathrm{def} \sim 2.0 N_\mathrm{merg} M_\bullet$ in our merger sequences, a value up to $\sim4$ times higher than in the binary + bulge experiments of \cite{Merritt2006}. The more observationally motivated 2D mass deficits of $2.2\lesssim M_\mathrm{def,2D}/M_\bullet\lesssim 3.1$ are in good agreement with the population of local cored ETGs \citep{Hopkins2010, Rusli2013}. The scaling of the 2D mass deficit $M_\mathrm{def,2D}$ as a function of SMBH mass $M_\bullet$ is steeper than the linear relation $M_\mathrm{def,2D}\propto M_\bullet$, again consistent with the observed mass and light deficits in the local core ETG population \citep{Rusli2013,Kormendy2009}.

The effect of core scouring is not strictly restricted within the influence radii of the central SMBHs as the SMBH binary can eject any star on an orbit with low enough angular momentum. Thus, global properties of ETGs such as their effective radii can be affected as well, connecting the small-scale post-Newtonian SMBH binary dynamics to the large-scale galactic structure. Extreme mass deficits of $>20\%$ of the total stellar mass caused by either overmassive SMBHs or a large number of merger generations can increase the effective radii $R_\mathrm{e}$ by up to a factor of $2$ due to scoured stars being ejected or re-deposited in the outer parts of the galaxies with low binding energies. We have presented a simple analytic scouring model which qualitatively explains the additional galaxy size growth due to SMBH scouring in the numerical simulations.

The shapes of our simulated merger remnants are mostly close to spherical or mildly axisymmetric with only one model having an axis ratio less than $b/a<0.9$. We attribute the shapes of the simulated merger remnants to the fraction of tube orbits compared to radial box orbits within the half-mass radius of the remnant.
As a consequence of the dominance of tube orbits, the central regions of the final simulated merger remnants are tangentially biased ($\beta_\mathrm{rot}<0$) within $r\lesssim2$--$3 R_\mathrm{b}$, consistent with the observed orbital anisotropies of local core galaxies \citep{Thomas2014,Mehrgan2019,Neureiter2023,deNicola2024}.

Our disky galaxy models (D-MA-BH) and spherical setups (S-MA-FROT-BH) that become fast-rotating due to a major merger orbit gradually evolve towards slow rotation ($\lambda_\mathrm{R}<0.1$) in repeated minor mergers. This is due to the fact that isotropic minor mergers bring in stellar mass and no net angular momentum. The models would be classified as slow rotators after $6$ and $8$ minor mergers, respectively. Only the disky models D-MI-BH without a major merger remain fast rotating after seven minor mergers.

Simulating the evolution of $z\gtrsim2$ compact ETGs into present-day massive, cored, slowly rotating ETGs self-consistently in cosmological zoom-in simulations with \ketju{} remains an important future goal. In cosmological zoom-in simulations, the hierarchical galaxy assembly provides the major and minor mergers naturally, without needing to specify the numbers, times and mass ratios of mergers by hand. The role of triple SMBH interactions occurring in cosmological simulations \citep{Mannerkoski2021} cannot be analyzed in idealized binary merger simulations, such as this study. In addition, \ketju{} has support for relativistic GW recoil kicks which we for simplicity did not include in this study. 

The first \ketju{} zoom-in simulations performed by \cite{Mannerkoski2021,Mannerkoski2022} did not specifically target compact ETGs and enabled \ketju{} integration only at $z\lesssim0.8$ and thus cannot address the early evolution ($0.8\lesssim z \lesssim 2.0$) of the nuclei of compact ETGs in mergers. Nevertheless, the stellar mass resolution and the stellar gravitational softening length of \cite{Mannerkoski2022} are within a factor of $\sim3$ of the parameter values in this study. Thus, besides increased computational costs due to initially very compact host galaxies of the targeted SMBHs there is no fundamental obstacle for studying the evolution of red nuggets in self-consistently cosmological zoom simulations with \ketju{}. Such simulations would help to disentangle the relationships of high-redshift red nuggets, local relic galaxies and intermediate-redshift relic candidates with varying degrees of relicness, further scrutinizing the role of minor and major mergers in the evolution of massive early-type galaxies and the corner-cases of galaxy evolution.

\section*{Data availability statement}
The data relevant to this article will be shared on reasonable request to the corresponding author.

\section*{Acknowledgments}
A. Rantala thanks Stavros Pastras and Natalia Lah\'en for their custom version of the velocity map program in the \textsc{python} package \texttt{pygad} \citep{Rottgers2020}. A. Rawlings acknowledges the support by the University of Helsinki Research Foundation and the support of the Academy of Finland grant
339127. TN acknowledges support from the Deutsche Forschungsgemeinschaft (DFG, German Research Foundation) under Germany's Excellence Strategy - EXC-2094 - 390783311 from the DFG Cluster of Excellence "ORIGINS". 
 PHJ acknowledges the support by the European Research Council via ERC Consolidator Grant KETJU (no. 818930) and the support of the Academy of Finland grant 339127.
The numerical simulations were performed using facilities hosted by the Max Planck Computing and Data Facility (MPCDF) in Garching, Germany.


\bibliographystyle{mnras}
\interlinepenalty=10000
\bibliography{references}


\appendix

\section{Major mergers orbits and the merger remnant rotation}\label{section: appendix-rotation}

\begin{figure}
\includegraphics[width=\columnwidth]{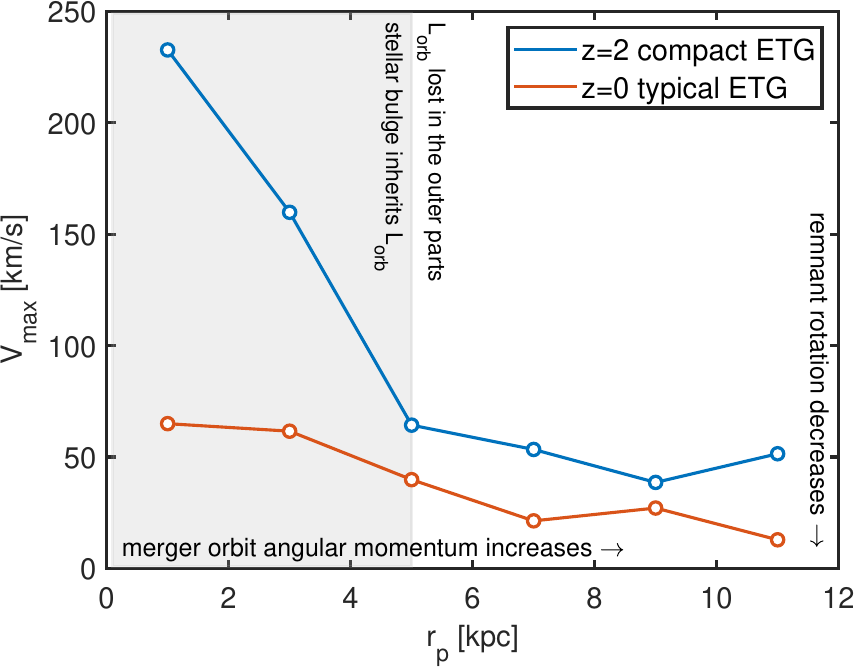}
\caption{The connection of the pericenter distance $r_\mathrm{p}$ of the galaxy merger orbit and the maximum rotation velocity $V_\mathrm{max}$ of the stellar component merger remnant. Larger $r_\mathrm{p}$ correspond to more initial angular momentum. However, for larger $r_\mathrm{p}$ the stellar component loses more of its angular momentum to the halo before the stellar bulges merge, resulting in less angular momentum in the stellar remnant, and thus lower values for $V_\mathrm{max}$. Compact progenitor ETGs result in more rapidly rotating remnants than more typical, extended ETGs. This is simply due to the conservation of angular momentum.}
\label{fig: orbit-Vrot}
\end{figure}

Consider a parabolic galaxy merger orbit with an initial separation of $r_\mathrm{sep}$ and a pericenter distance $r_\mathrm{p}$. The two galaxies in this example consist of an initially non-rotating, spherical stellar bulge and a dark matter halo. For separations smaller than the virial radii of the halos the orbital velocity is typically scaled with $M(<r_\mathrm{sep})/M_\mathrm{tot}$ rendering the orbit almost parabolic. Larger values for the pericenter distance $r_\mathrm{p}$ correspond to more orbital angular momentum $\norm{\vect{L_\star}}$ for the stellar bulges. As the stellar bulges do not initially rotate, all stellar angular momentum is initially in the orbit.

For small $r_\mathrm{p}$, the galaxy merger process is swift and the stellar bulges merge rapidly. The merged stellar remnant inherits most of the angular momentum of the orbit, the rest is transferred to the dark matter component. However, if $r_\mathrm{p}$ is large, the stellar bulges orbit for extended times in the halos, transferring their angular momentum to the halo particles and eventually merging from almost radial orbits with little angular momentum. The merged stellar remnant inherits whatever little angular momentum is left at this phase. We show the maximum remnant rotation velocities for galaxy mergers with different orbits in Fig. \ref{fig: orbit-Vrot}. We display six different values for initial angular momentum (or $r_\mathrm{p}$) and two different galaxy models, a compact ETG ($r_\mathrm{h}=1$ kpc) and a typical ETG with $r_\mathrm{h}=5$ kpc. Larger $r_\mathrm{p}$ corresponds to less rotation in the merger remnant.

An order-of-magnitude estimate for the rotation velocity of the stellar component of the galaxy after the merger can be estimated as $V_\mathrm{max} \sim \norm{\vect{L_\star}} M_\mathrm{\star}^\mathrm{-1} r_\mathrm{h}^\mathrm{-1}$. Thus, the compact ETG setup results in overall higher remnant rotation velocities up to $V_\mathrm{max} \sim 235$ km/s. The typical, more extended ETGs can only acquire rotation velocities of up to $V_\mathrm{max} \sim 60$ km/s.


\section{Additional idealized simulation setups} \label{section: appendix-sims}

\subsection{SMBH binary eccentricity and core scouring}\label{section: appendix-sims-gwecc}

\begin{figure}
\includegraphics[width=\columnwidth]{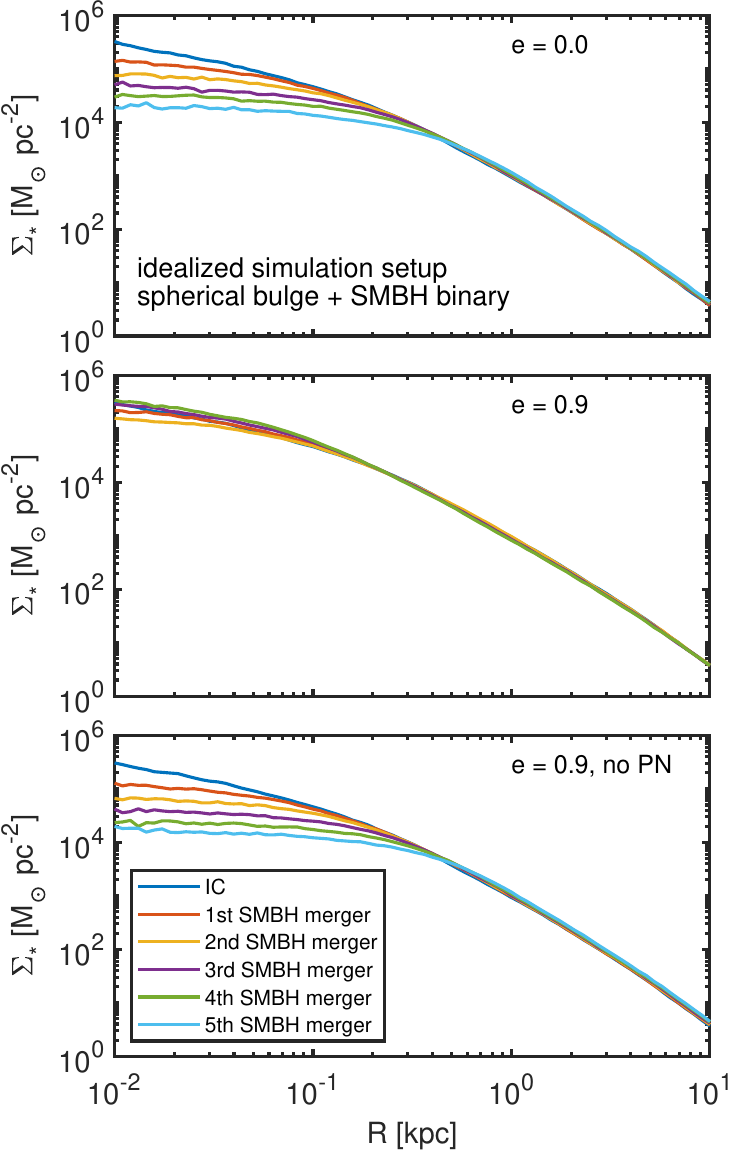}
\caption{Idealized core scouring simulations of a stellar bulge and embedded unequal-mass SMBH binaries. The different lines indicate the initial profile and five subsequent SMBH merger generations. The top panel shows experiments I-ECC-0-PN with consecutive initially circular SMBH binaries ($e=0.0$) in which the binaries scour a flat, low-density core. In the middle panel (runs I-ECC-9-PN) the initial binary eccentricity is $e=0.9$ and the binaries rapidly merge via GW emission without scouring a flat core, similarly to the sequence S-MI-BH in this study. The bottom panel shows a numerical experiment I-ECC-9-noPN in which the PN equations of motion are disabled and the eccentric binaries ($e=0.9$) also can scour a core.}
\label{fig: idealized-scouring}
\end{figure}

We perform an idealized test simulation set following \cite{Merritt2006} in order to assess the effect of the initial SMBH binary eccentricity on the core scouring. We setup an isolated Hernquist galaxy model with $M_\star=\msol{1.2\times10^{11}}$ and $r_\mathrm{h}=1.2$ kpc. The dark halo mass is $M_\mathrm{dm}=\msol{1.2\times10^{12}}$ with $f_\mathrm{dm}=10\%$ within the stellar $r_\mathrm{h}$. The initial SMBH mass is $M_\bullet=\msol{2\times10^9}$. The initial SMBH is placed in a binary at the center of the stellar bulge with a secondary SMBH with $M_\mathrm{\bullet,2}=\msol{4\times10^8}$. The semi-major axis of the binary is $a=5$ pc.

\begin{table}
    \centering
    \begin{tabular}{lccl}
        \hline
        sequence & $e$ & PN & run time\\
        \hline
        I-ECC-0-PN & $0.0$ & $\checkmark$ & until merger at $t=t_\mathrm{GW,e0PN}$\\
        I-ECC-9-PN & $0.9$ & $\checkmark$ & until merger\\
        I-ECC-9-noPN & $0.9$ & $\times$ & until $t=t_\mathrm{GW,e0PN}$ \\
         \hline
    \end{tabular}
    \caption{The idealized core scouring simulation set varying the initial SMBH binary eccentricity and the use of PN equations of motion.}
    \label{table: merritt-runs}
\end{table}

We run in total three new idealized simulation sequences varying the initial SMBH binary eccentricity and with and without PN equations of motion. The sequences and their main properties are listed in Table \ref{table: merritt-runs}. The sets I-ECC-0-PN and I-ECC-9-PN include the post-Newtonian terms and have initial binary eccentricities of $e=0.0$ and $e=0.9$, respectively. The simulations are run until the GW merger after which a new binary with the merger remnant and a new secondary with $M_\mathrm{\bullet,2}=\msol{4\times10^8}$ is initialized. The third and final setup I-ECC-9-noPN has a high initial binary eccentricity $e=0.9$ but does not include PN equations of motion. The simulation run times in this setup are determined by the binary merger times in the circular PN setup I-ECC-0-PN.

We show the results of the idealized binary scouring experiments in Fig. \ref{fig: idealized-scouring}. Circular binaries of the sequence I-ECC-0-PN scour flat, extended, low-density cores with mass deficits consistent with \cite{Merritt2006}. In the eccentric ($e=0.9$) binary merger sequence I-ECC-9-PN most of the binary orbital energy is lost to GW emission and not transferred to the stellar component via three-body interactions. Thus, the SMBH binaries can merge without scouring a core. This is what occurs in the sequence S-MA-BH of this study as well. If PN terms are disabled as in the setup I-ECC-9-noPN, the binaries scour a flat low-density core comparable to the circular setup I-ECC-0-PN. We conclude that general relativistic effects (together with initial binary orbits) can affect the central structure of the galaxy on a spatial scale of hundreds of parsecs.

\subsection{Scouring-driven growth of $R_\mathrm{e}$: merging the SMBH binary by hand}\label{section: appendix-sims-byhand}

\begin{figure}
\includegraphics[width=\columnwidth]{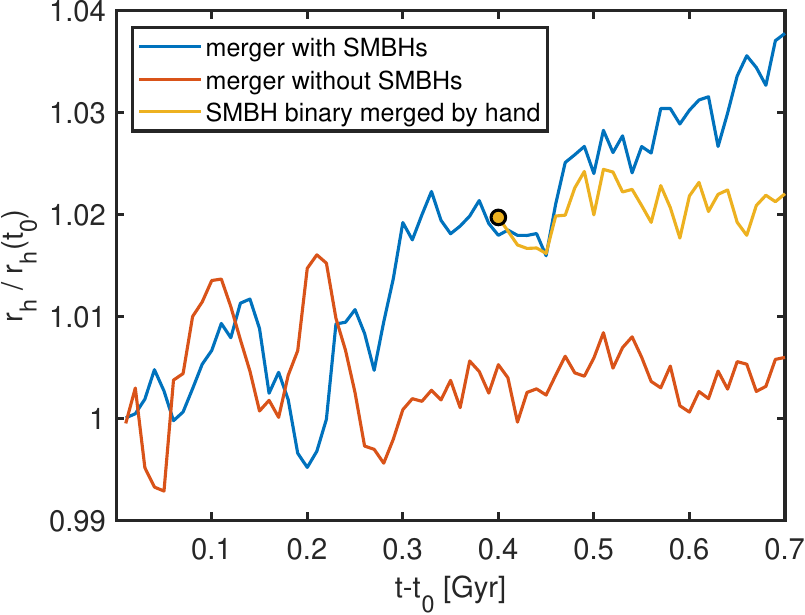}
\caption{The experiment to confirm the role of SMBH core scouring in driving the additional size growth of galaxies in the aftermath of galaxy mergers. Without SMBHs, there is no additional size growth. If the SMBH binary is merged by hand (filled circle), the galaxy size growth is immediately stalled.}
\label{fig: merged-by-hand}
\end{figure}

We test our hypothesis presented in Section \ref{section: scouring-rh} that core scouring with large mass deficits can increase the effective radius $R_\mathrm{e}$ of the galaxies. We run two additional galaxy merger test simulations with spherical progenitor properties ($M_\star=\msol{10^{11}}$, $M_\bullet=\msol{2\times10^9}$, $r_\mathrm{h}=5$ kpc) as in the typical ETG setup of Appendix \ref{section: appendix-rotation}, one with and another run without SMBHs. As with the merger sequences of this study, the half-mass radius $r_\mathrm{h}$ is larger in the model with SMBHs. 

We present $r_\mathrm{h}$ as a function of time in the test simulations in Fig. \ref{fig: merged-by-hand}. The SMBHs form a bound binary around time $t_\mathrm{0}$. Without SMBHs, the half-mass radius of the merger remnant remains relatively constant after the stellar bulges have merged. On the other hand, with SMBHs the $r_\mathrm{h}$ grows close to linearly as a function of time. We run a third simulation based on the run with SMBHs in which we merge the SMBH binary by hand when $r_\mathrm{h}/r_\mathrm{h}(t_\mathrm{0})=1.02$. In this simulation, the galaxy size growth abruptly stops, confirming the role of the SMBH core scouring in driving the additional merger remnant size growth in the aftermath of galaxy mergers.

\section{A simple analytic model for size evolution due to SMBH
scouring: the Dehnen profile} \label{section: appendix-gamma}

\begin{figure}
\includegraphics[width=\columnwidth]{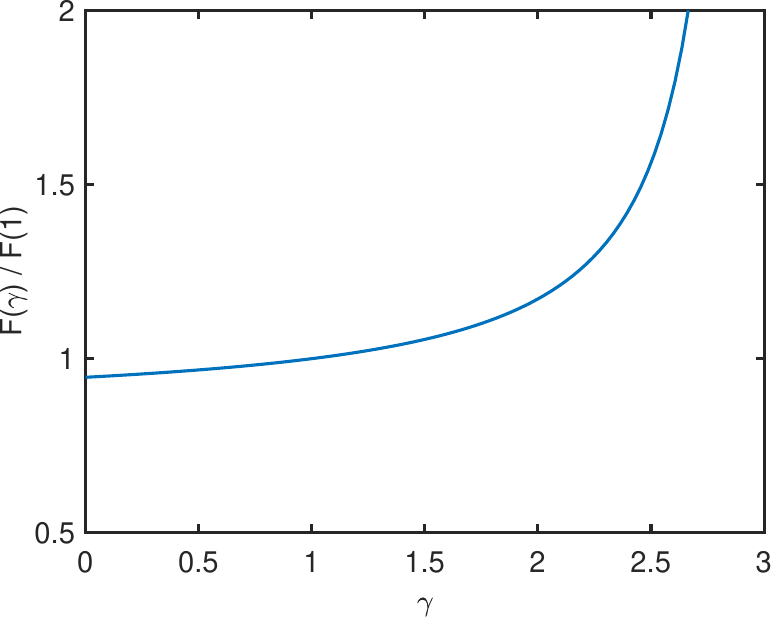}
\caption{The function $F(\gamma)$ of Eq. \eqref{eq: rh-increase-dehnen} normalized to the Hernquist ($\gamma=1$) value of $F(1)$.}
\label{fig: gamma}
\end{figure}

For the \cite{Dehnen1993} galaxy model with a central density slope in the range of $0\leq \gamma < 3$, the relative size change due to SMBH core scouring from Eq. \eqref{eq: rh-increase} can be expressed as
\begin{equation}\label{eq: rh-increase-dehnen}
\begin{split}
\frac{\Delta R_\mathrm{e}}{R_\mathrm{e}} &= F(\gamma) \frac{\eta M_\mathrm{def}} {M_\star}\\
F(\gamma) &= \frac{(x+1)^{4-\gamma}}{(3-\gamma) x^{3-\gamma}}\\
\end{split}
\end{equation}
in which the argument $x=x(\gamma)$ is directly obtained from the definition of the half-mass radius for the Dehnen profile:
\begin{equation}
x = \frac{r_\mathrm{h}}{a} = \left[ 2^{1/(3-\gamma)} -1\right]^{-1}.
\end{equation}
For the Hernquist profile $F(1)=2+\sqrt{2}$ as in Eq. \eqref{eq: rh-increase-hernquist}. The parameter $F(\gamma)$ depends only mildly on $\gamma$ for typical ETG central slopes of $\gamma\lesssim2.5$. We demonstrate this fact in Fig. \ref{fig: gamma}. Centrally flat ($\gamma=0$) and \cite{Jaffe1983} ($\gamma=2$) profiles differ from the Hernquist value for $F(\gamma)$ by only $\sim5\%$ ($F(0)/F(1)\sim0.95$) and  $\sim17\%$ ($F(2)/F(1)\sim1.17$). Intuitively, the scouring size increase is sensitive to the slope of the mass profile at $r_\mathrm{h}$ where Dehnen profiles with $\gamma\lesssim2.0$ do not considerably differ from each other. For very steep cusps with $\gamma=2.5$ we have $F(2.5)/F(1)\sim1.56$.


\bsp	
\label{lastpage}
\end{document}